\documentclass[dvipdfmx]{article}
\usepackage{tikz}

\begin{document}

\begin{center}

{\large 
Honeycomb lattice Kitaev model 
with Wen-Toric-code interactions, \\
and anyon excitations.
}
\vspace{0.6cm}


Kazuhiko Minami
\vspace{0.6cm}

Graduate School of Mathematics, Nagoya University, \\
Furo-cho, Chikusa-ku, Nagoya, Aichi, 464-8602, JAPAN.

\end{center}

\begin{abstract}
The honeycomb lattice Kitaev model ${\cal H}_{K}$ 
with two kinds of Wen-Toric-code four-body interactions ${\cal H}_{WT}$ 
is investigated exactly  
using a new fermionization method,  
and the ground state phase diagram is obtained. 
Six kinds of three-body interactions are also considered. 
A Hamiltonian equivalent to the honeycomb lattice Kitaev model 
is also introduced. 
The fermionization method is generalized to two-dimensional systems, 
and the two-dimensional Jordan-Wigner transformation is obtained 
as a special case of this formula. 
The model ${\cal H}_{K}+{\cal H}_{WT}$ is symmetric 
in four-dimensional space of coupling constants, 
and the anyon type excitations appear in each phase. 
\end{abstract}



\section{Introduction}\label{introduction}

Recently, a new fermionization formula was introduced 
\cite{16Minami}, 
in which solvable Hamiltonians 
and the transformations to diagonalize them 
can be obtained simultaneously. 
The one-dimensional transverse Ising model, XY model, cluster model, 
the two-dimensional square lattice Ising model,  
and an infinite number of unsolved models were diagonalized by this formula. 
The Jordan-Wigner transformation is obtained as a special case of this tratment.\cite{16Minami}\cite{17Minami} 

The formula is summarized as follows: 
Let us consider a series of operators $\{\eta_j\}$ $(j=1, 2, \ldots, M)$. 
The operators $\eta_j$ and $\eta_k$ are called 'adjacent' 
when $(j, k)=(j, j+1)$ $(1\leq j\leq M-1)$, or $(j, k)=(M, 1)$. 
If the operators $\eta_j$ satisfy the relations 
\begin{eqnarray}
\eta_{j}\eta_{k}
=
\left\{
\begin{array}{cl}
1 & j=k \\
-\eta_{k}\eta_{j} & \eta_{j}\: {\rm and}\: \eta_{k} \:{\rm are}\: {\rm adjacent} \\
\eta_{k}\eta_{j} &{\rm otherwise,} \\
\end{array}
\right.
\label{cond}
\end{eqnarray}
then we can introduce a solvable Hamiltonian 
\begin{eqnarray}
-\beta{\cal H}
=
\sum_{j=1}^{M}K_j \eta_j,
\label{hamKjetaj}
\end{eqnarray}
which can be mapped to the free fermion system by the transformation 
\begin{eqnarray}
\varphi_j
=
\frac{1}{\sqrt{2}}
e^{i\frac{\pi}{2}(j-1)}
\eta_0
\eta_1
\eta_2
\cdots
\eta_j
\hspace{0.6cm}
(0\leq j\leq M), 
\label{transmain}
\end{eqnarray}
where $\eta_0$ is an initial operator satisfying 
$\eta_0^2=-1$,  $\eta_0\eta_1=-\eta_1\eta_0$,  
and $\eta_0\eta_k=\eta_k\eta_0$ $(2\leq k\leq M)$. 
The operators $\varphi_j$ satisfy 
$(-2i)\varphi_{j}\varphi_{j+1}=\eta_{j+1}$,
and  
\begin{eqnarray}
\{\varphi_j, \varphi_k\}
=
\varphi_j\varphi_k+\varphi_k\varphi_j
=
\delta_{jk}.
\label{anticom}
\end{eqnarray}
Hence the Hamiltonian (\ref{hamKjetaj}) is expressed 
as a sum of two-body products of the fermion operators $\varphi_{j}$, 
and can be diagonalized. 

The transformation (\ref{transmain}) is automatically generated 
from the series of operators $\{\eta_j\}$, 
and only the algebraic relations (\ref{cond}), 
together with the translational invariance, 
are needed to obtain the free energy. 
This procedure can be applied to any systems 
written by the operators  
that satisfy (\ref{cond}). 

The one-dimensional XY model, 
and its generalizations\cite{71Suzuki}  
can be solved by this formula. 
In these cases, the transformation (\ref{transmain}) 
results in the Jordan-Wigner transformation. 

The one-dimensional cluster model 
with the next-nearest-neighbor interaction 
\begin{eqnarray}
-\beta{\cal H}
=
\sum_{j=1}^{N}[ 
K_{1}\sigma^{x}_{j}\sigma^{z}_{j+1}\sigma^{x}_{j+2}
+K_{2}\sigma^{x}_{j+1} \: 1_{j+2}\sigma^{x}_{j+3}
]
\label{HamClusterX1X}
\end{eqnarray}
cannot be diagonalized by the Jordan-Wigner transformation.  
This model, however, can be decoupled into 
${\cal H}={\cal H}_{even}+{\cal H}_{odd}$, 
where $j=$even in ${\cal H}_{even}$, and $j=$odd in ${\cal H}_{odd}$, 
respectively. 
They satisfy $[{\cal H}_{even}, {\cal H}_{odd}]=0$, 
and ${\cal H}_{even}$, for example, is obtained
from a series of operators 
\begin{eqnarray}
\eta_{2j-1}=\sigma^{x}_{2j-1}\sigma^{z}_{2j}\sigma^{x}_{2j+1},  
\hspace{0.6cm} 
\eta_{2j}=\sigma^{x}_{2j} \: 1_{2j+1}\sigma^{x}_{2j+2}, 
\label{seriesClusterNnn}
\end{eqnarray}
which satisfy (\ref{cond}).  
In this case, the transformation (\ref{transmain}) becomes 
\begin{eqnarray}
\varphi_{2j}
&=&
\frac{1}{\sqrt{2}}
(\prod_{\nu=1}^{j}1_{2\nu-1}\sigma_{2\nu}^z)
\sigma_{2j+1}^x\sigma_{2j+2}^x,
\nonumber
\\
\varphi_{2j+1}
&=&
\frac{1}{\sqrt{2}}
(\prod_{\nu=1}^{j}1_{2\nu-1}\sigma_{2\nu}^z)
1_{2j+1}\sigma_{2j+2}^y\sigma_{2j+3}^x
\nonumber\\
&&\hspace{2.8cm}
(j=0, 1, 2, 3, \ldots),
\label{trans3}
\end{eqnarray}
which is apparently different from the Jordan-Wigner transformation, 
and the Hamiltonian (\ref{HamClusterX1X}) is diagonalized 
through this formula.\cite{17Minami}\cite{18Yanagihara}

In this paper, this formula is applied to two-dimensional systems. 
The transformation (\ref{transmain}) is generally formulated for the square lattice. 
The Hamiltonian 
\begin{eqnarray}
{\cal H}
=
{\cal H}_{K}+{\cal H}_{3}+{\cal H}_{WT}
\label{totHamiltonian}
\end{eqnarray}
is transformed to the fermion system,  
and the ground state of ${\cal H}_{K}+{\cal H}_{WT}$ is exactly specified;    
here 
${\cal H}_{K}$ denotes the Hamiltonian of the honeycomb lattice Kitaev model,  
${\cal H}_{3}$ consists of the six kinds of three-body interactions,  
${\cal H}_{WT}$ denotes the Hamiltonian of the Wen model 
which is equivalent to the Kitaev toric-code model. 
A Hamiltonian, 
which consists of the cluster-type chains coupled by the Ising interactions, 
is obtained as an system equivalent to ${\cal H}_{K}$. 
The ground-state phase diagram of ${\cal H}_{K}+{\cal H}_{WT}$ is obtained exactly, 
and it is depicted that the phase structure of gapped phases and gapless phases 
change with the rates of the interactions. 
The symmetry of the system is investigated, 
and it is derived that 
the system is symmetric in four-dimensional space of coupling constants. 
The anyon excitations exist in each phase.

In section \ref{Hamiltonian}, 
the honeycomb lattice Kitaev model 
and the three-body interactions are introduced. 
The Wen model is also introduced 
and the relation with the Kitaev toric-code model is considered. 
In section \ref{transformation}, 
the transformation (\ref{transmain}) is generally formulated 
for the two-dimensional square lattice.  
A specific series of operators is then introduced 
to obtain and diagonalize the Hamiltonian (\ref{totHamiltonian}).  
The transformation (\ref{transmain}) in this case 
is found to be the two-dimensional Jordan-Wigner transformation. 
In section \ref{OperatorsAndInteractions}, 
the interactions are expressed by Majorana fermion operators. 
Operators that commute with the Hamiltonian are also introduced. 
In section \ref{Generalizations}, 
the series of operators is rearranged, 
and in section \ref{EquivalentHamiltonian}, 
a Hamiltonian equivalent to the honeycomb lattice Kitaev model is introduced. 
In section \ref{Diagonalization}, 
${\cal H}_{K}+{\cal H}_{WT}$ is diagonalized in a subspace 
containing one of the ground states. 
In section \ref{GaplessCondition}, 
the gapless condition is derived, 
and in section \ref{SymmetriesAndDiscussions}, 
the ground state phase diagram is obtained. 
Symmetries of the model is investigated 
and it is pointed out that 
the anyon excitations appear in each phase.

\section{Hamiltonian}\label{Hamiltonian}

\begin{figure}
\includegraphics[scale=1.0]{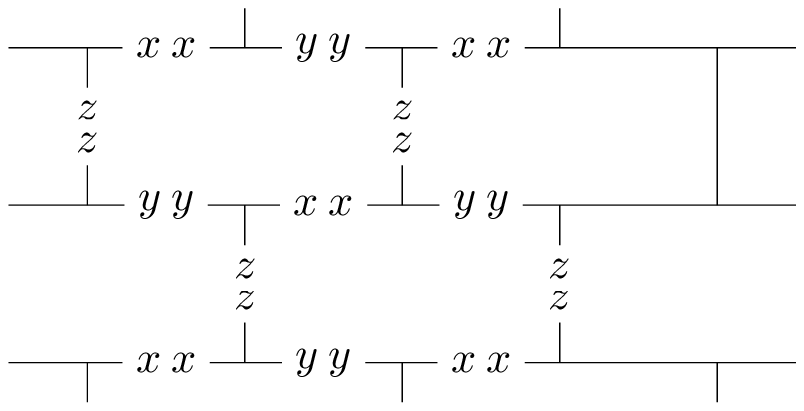}
\caption{\label{fig:BrickWallLattice} Honeycomb lattice Kitaev model on the brick-wall lattice. }
\end{figure}

\begin{figure}
\includegraphics{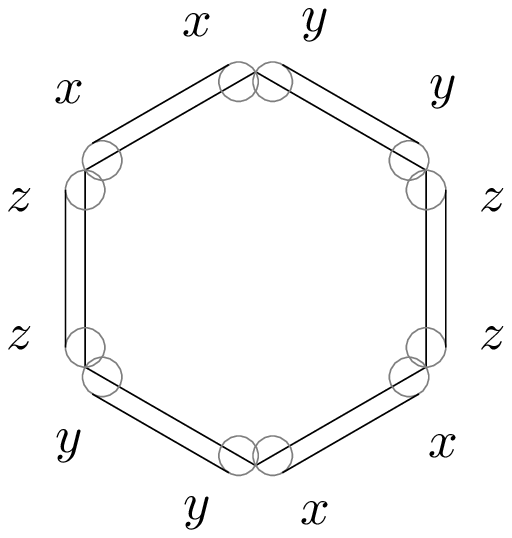}
\caption{\label{fig:KitaevInteractions} Two-body interactions of the honeycomb lattice Kitaev model. }
\end{figure}

Let us consider the brick-wall lattice, shown in Fig.\ref{fig:BrickWallLattice} 
(see also Fig.\ref{fig:Operators}), 
with the interactions 
\begin{eqnarray}
-\beta{\cal H}_{K}
&=&
\sum_{{l=1}}^{M_{2}}
\Big[
K_{x}\sum_{j=odd}^{M_{1}-1}
\sigma^{x}_{j\: l}
\sigma^{x}_{j+1\: l}
+
K_{y}\sum_{j=odd}^{M_{1}-1}
\sigma^{y}_{j-1\: l}
\sigma^{y}_{j\: l}
\nonumber\\
&&
\hspace{4.0cm}+
K_{z}\sum_{j=odd}^{M_{1}-1}
\sigma^{z}_{j-1\: l}
\sigma^{z}_{j\: l+1}
\Big],
\label{HamiltonianKitaev}
\end{eqnarray}
where $M_{1}$ is even and the summation is taken over all odd $j$. 
Hamiltonian (\ref{HamiltonianKitaev}) is the Kitaev model 
on the honeycomb lattice 
shown in Fig.\ref{fig:KitaevInteractions}. 
The Kitaev model is introduced in \cite{06Kitaev},  
in which the ground state is specified, 
the phase diagram is obtained, 
and abelian anyon excitations in gapped phases, 
and non-abelian anyon excitations in gapless phases 
are found.

\begin{figure}
\includegraphics{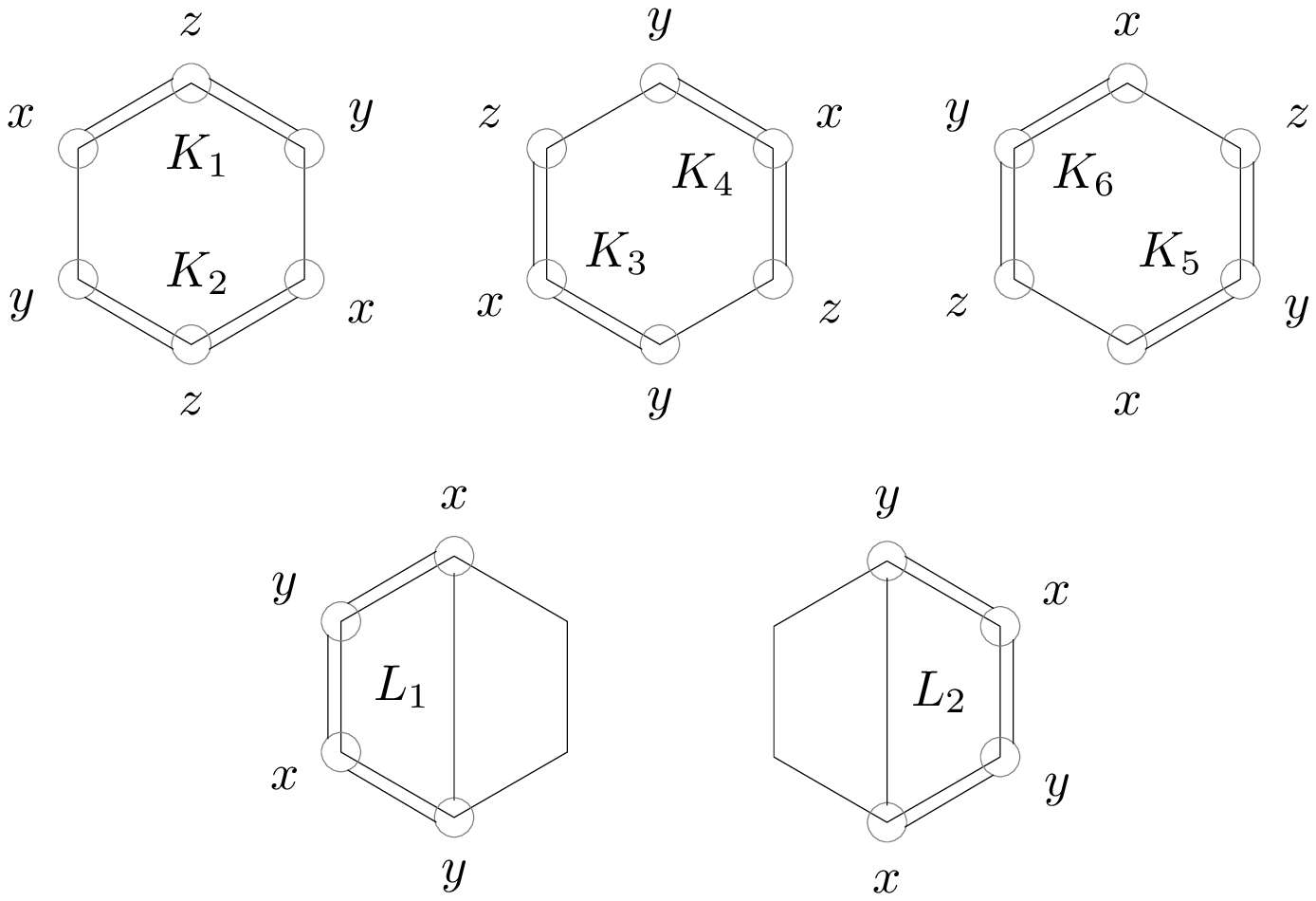}
\caption{\label{fig:3bodyAnd4bodyInteractions}Three-body and Wen-Toric-code four-body interactions.}
\end{figure}

We will also introduce six kinds of three-body interactions 
shown in Fig.\ref{fig:3bodyAnd4bodyInteractions} as 
\begin{eqnarray}
&&
-\beta{\cal H}_{3}
=
\sum_{{l=1}}^{M_{2}}
\sum_{j=odd}^{M_{1}-1}
\Big[\:
K_{1}
\sigma^{x}_{j\: l}
\sigma^{z}_{j+1\: l}
\sigma^{y}_{j+2\: l}
+
K_{2}
\sigma^{y}_{j-1\: l}
\sigma^{z}_{j\: l}
\sigma^{x}_{j+1\: l}
\nonumber\\
&&
\hspace{2.8cm}
+
K_{3}
\sigma^{y}_{j\: l}
\sigma^{x}_{j-1\: l}
\sigma^{z}_{j\: l+1}
+
K_{4}
\sigma^{z}_{j-1\: l}
\sigma^{x}_{j\: l+1}
\sigma^{y}_{j-1\: l+1}
\nonumber\\
&&
\hspace{2.6cm}
+
K_{5}
\sigma^{x}_{j-2\: l}
\sigma^{y}_{j-1\: l}
\sigma^{z}_{j\: l+1}
+
K_{6}
\sigma^{z}_{j-1\: l}
\sigma^{y}_{j\: l+1}
\sigma^{x}_{j+1\: l+1}
\:\Big].
\label{Hamiltonian3}
\end{eqnarray}

These three-body interactions are already investigated by several authors. 
Lee et al.\cite{07LeeZhangXiang} and Shi et al.\cite{09ShiYuYouNori} 
introduced the interactions $K_{1}$ and $K_{2}$, and 
Yu and Wang\cite{08YuWang} and Yu\cite{08Yu} 
introduced from $K_{3}$ to $K_{6}$.
Yu\cite{08Yu} also introduced 
various kinds of four-body and six-body interactions.

Let us here consider the Wen model\cite{03Wen}. 
The Hamiltonian is given by 
\begin{eqnarray}
&&
-\beta{\cal H}_{WT}
=
\sum_{{l=1}}^{M_{2}}
\Big[
L_{1}\sum_{j=odd}^{M_{1}-1}
\sigma^{y}_{j\: l+1}
\sigma^{x}_{j+1\: l+1}
\sigma^{y}_{j\: l}
\sigma^{x}_{j-1\: l}
\nonumber\\
&&
\hspace{3.6cm}+
L_{2}\sum_{j=odd}^{M_{1}-1}
\sigma^{y}_{j+1\: l+1}
\sigma^{x}_{j+2\: l+1}
\sigma^{y}_{j+1\: l}
\sigma^{x}_{j\: l}
\Big], 
\label{HamiltonianWen}
\end{eqnarray}
and the interactions are shown in Fig.\ref{fig:3bodyAnd4bodyInteractions}. 
Wen originally introduced the case $L_{1}=L_{2}$, 
and investigated the ground state quantum orders. 

\begin{figure}
\includegraphics{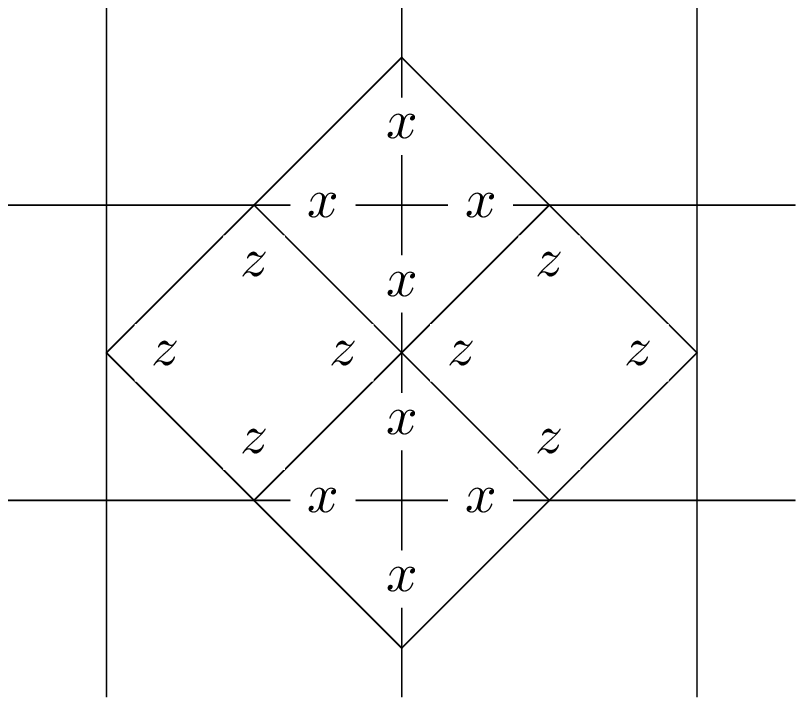}
\caption{
\label{fig.TorcCodeInteractions} 
Kitaev toric-code model. 
}
\end{figure}

We will also consider the Kitaev toric-code model\cite{03Kitaev}, 
which consists of two types of interactions, 
as shown in Fig.\ref{fig.TorcCodeInteractions}. 
The spin variables are located on each edge. 
Let us consider the spins on the vertical edges 
and consider a canonical transformation 
\begin{eqnarray*}
\sigma^{x}_{jl}\mapsto \sigma^{z}_{jl},
\hspace{0.4cm}
\sigma^{z}_{jl}\mapsto \sigma^{x}_{jl},
\hspace{0.4cm}
\sigma^{y}_{jl}\mapsto -\sigma^{y}_{jl}, 
\end{eqnarray*}
and next another transformation of all spins
\begin{eqnarray*}
\sigma^{x}_{jl}\mapsto \sigma^{y}_{jl},
\hspace{0.4cm}
\sigma^{y}_{jl}\mapsto \sigma^{z}_{jl},
\hspace{0.4cm}
\sigma^{z}_{jl}\mapsto \sigma^{x}_{jl}, 
\end{eqnarray*}
then we find the Wen model on the square lattice 
rotated by $\pi/4$ from the original square lattice. 
Thus these two models are in this sense equivalent 
(see also sec.7.2 of \cite{06Kitaev}). 

In the case of the honeycomb lattice Kitaev model 
having only two-body interactions, 
the Hamiltonian commutes with the following operators 
associated to each hexagon 
\begin{eqnarray}
W_{jl}
=
\sigma^{x}_{j-1\: l}\sigma^{y}_{j\: l+1}\sigma^{z}_{j+1\: l+1}
\sigma^{x}_{j+2\: l+1}\sigma^{y}_{j+1\: l}\sigma^{z}_{j\: l}.
\label{Wjl}
\end{eqnarray}
Each $W_{jl}$ has the eigenvalues $w_{jl}=\pm 1$. 
It is easy to demonstrate that 
the Hamiltonian with the three-body interactions (\ref{Hamiltonian3}) 
and with the Wen-Toric-code four-body interactions (\ref{HamiltonianWen})
also commute with all $W_{jl}$.  
The eigenstates of the Hamiltonian 
may thus be labelled by the set of eigenvalues of $W_{jl}$, 
and the total Hilbert space is divided into subspaces 
labelled by $\{w_{jl}\}$. 

It should be noted that 
the interactions $L_{1}$ and $L_{2}$ are not independent. 
When we consider the product of the four-body terms, 
we find the following relation 
\begin{eqnarray*}
&&
(L_{1}
\sigma^{y}_{j\: l+1}
\sigma^{x}_{j+1\: l+1}
\sigma^{y}_{j\: l}
\sigma^{x}_{j-1\: l})
(L_{2}
\sigma^{y}_{j+1\: l+1}
\sigma^{x}_{j+2\: l+1}
\sigma^{y}_{j+1\: l}
\sigma^{x}_{j\: l})
\\
&&
=
L_{1}
L_{2}
\sigma^{y}_{j\: l+1}
(\sigma^{x}_{j+1\: l+1}
\sigma^{y}_{j+1\: l+1})
\sigma^{x}_{j+2\: l+1}
\sigma^{y}_{j+1\: l}
(\sigma^{y}_{j\: l}
\sigma^{x}_{j\: l})
\sigma^{x}_{j-1\: l}
\\
&&
=
L_{1}L_{2}
\sigma^{y}_{j\: l+1}
\sigma^{z}_{j+1\: l+1}
\sigma^{x}_{j+2\: l+1}
\sigma^{y}_{j+1\: l}
\sigma^{z}_{j\: l}
\sigma^{x}_{j-1\: l}
\\
&&
=
L_{1}L_{2}
W_{jl}.
\end{eqnarray*}

\section{Transformation}\label{transformation}


We will generalize (\ref{transmain}), 
and formulate the fermionization transformation for the two-dimensional lattice. 
Let us introduce operators $\eta_{kl}$ on each row $l$. 
The operators $\eta_{kl}$ with fixed $l$ satisfy the condition (\ref{cond}), 
and $\eta_{kl}$ on different row $l$ commute with each other. 
The series of operators on the first row is 
\begin{eqnarray*}
\eta_{11}, 
\hspace{0.3cm}
\eta_{21}, 
\hspace{0.3cm}
\ldots,
\hspace{0.3cm}
\eta_{M1} 
\end{eqnarray*}
with an initial operator $\eta_{01}$. 
The transformation is introduced as  
\begin{eqnarray}
\varphi_{01}
=
\frac{1}{\sqrt{2}}e^{i\frac{\pi}{2}(0-1)}
\eta_{01},
\hspace{0.6cm}
{\rm and}  
\hspace{0.6cm}
\varphi_{j 1}
=
\frac{1}{\sqrt{2}}e^{i\frac{\pi}{2}(j-1)}
\eta_{01}\eta_{11}\cdots\eta_{j 1},
\label{trans1row}
\end{eqnarray}
where $j=1, 2, \ldots, M$. 
From (\ref{trans1row}) we obtain 
\begin{eqnarray*}
\varphi_{j 1}\varphi_{j+1\: 1}
=
\frac{i}{2}\eta_{j+1\: 1}.
\end{eqnarray*}
At the end of the first row, we find 
\begin{eqnarray*}
\varphi_{M 1}\varphi_{1 1}
&=&
\Big(
\frac{1}{\sqrt{2}}i^{M-1}
\eta_{01}\eta_{11}\cdots\eta_{M 1}
\Big)
\Big(
\frac{1}{\sqrt{2}}i^{1-1}
\eta_{01}\eta_{11}
\Big)
\nonumber\\
&=&
i^{M-1}
(\eta_{11}\cdots\eta_{M 1})
\frac{1}{2}\eta_{11}
\nonumber\\
&=&
(-1)\: i^{M}\eta_{11}\cdots\eta_{M 1}\cdot \frac{i}{2}\eta_{11}.
\end{eqnarray*}
The operator $(-1)\: i^{M}\eta_{11}\cdots\eta_{M 1}$ 
commute with the Hamiltonian (\ref{hamKjetaj}), 
is hermitian and has the eigenvalues $\pm 1$. 
The Hilbert space is divided into two subspaces 
corresponding to the eigenvalues $+1$ and $-1$. 
We assume the periodic boundary condition for $\eta_{j1}$, 
and thus introduce the boundary condition for $\varphi_{j 1}$ as 
\begin{eqnarray}
\varphi_{M+1\: 1}
=
\left\{
\begin{array}{cl}
+\varphi_{1 1}, & (-1)\: i^{M}\eta_{11}\cdots\eta_{M 1}=+1 \\
-\varphi_{1 1}, & (-1)\: i^{M}\eta_{11}\cdots\eta_{M 1}=-1, 
\end{array}
\right.
\label{boundcond1strow}
\end{eqnarray}
in each eigenspace.


Next, let us consider the transformation for the second row. 
We will introduce the following factor that comes from the first row as 
\begin{eqnarray*}
H(1)
=
(-1)\: i^{M}\eta_{11}\cdots\eta_{M 1}.
\end{eqnarray*}
Then the transformation for the second row is defined as 
\begin{eqnarray}
\varphi_{j 2}
=
H(1)
\frac{1}{\sqrt{2}}e^{i\frac{\pi}{2}(j-1)}
\eta_{02}\eta_{12}\cdots\eta_{j 2}.
\label{trans2row}
\end{eqnarray}
The transformation (\ref{trans2row}) is schematically written as 
\begin{eqnarray}
\varphi_{j 2}
=
\frac{1}{\sqrt{2}}e^{i\frac{\pi}{2}(j-1)}\times 
\begin{array}{l}
\hspace{0.6cm}
\eta_{02}\cdot \eta_{12}\eta_{22}\cdots\eta_{j2} \\
(-1)\: i^{M}\eta_{11}\eta_{21}\cdots\eta_{j1}\cdots\eta_{M 1}. 
\end{array}
\label{trans2ndrow}
\end{eqnarray}
Note that $\eta_{02}$ is introduced in (\ref{trans2ndrow}), 
though $\eta_{01}$ is not introduced in $H(1)$.  
The boundary condition for the second row 
is obtained from (\ref{boundcond1strow}) 
replacing $\varphi_{j 1}$ by $\varphi_{j 2}$, and $\eta_{j 1}$ by $\eta_{j 2}$. 
There is no boundary in the first row.

Generally for the $l$-th row, the transformation is defined as 
\begin{eqnarray}
\varphi_{j l}
=
\Big(
\prod_{r=1}^{l-1}H(r)
\Big)
\frac{1}{\sqrt{2}}e^{i\frac{\pi}{2}(j-1)}
\eta_{0l}\eta_{1l}\cdots\eta_{j l},
\label{transgenrow}
\end{eqnarray}
where 
\begin{eqnarray*}
H(r)
=
(-1)\: i^{M}\eta_{1r}\cdots\eta_{M r}. 
\end{eqnarray*}
From (\ref{transgenrow}) we obtain 
\begin{eqnarray*}
\varphi_{j l}\varphi_{j+1\: l}
=
\frac{i}{2}\eta_{j+1\: l}.
\end{eqnarray*}
The boundary condition for the l-th row 
is obtained from (\ref{boundcond1strow}) 
replacing $\varphi_{j 1}$ by $\varphi_{j l}$, and $\eta_{j 1}$ by $\eta_{j l}$. 
It is easy to convince from (\ref{transgenrow}) that 
\begin{eqnarray*}
\{\varphi_{j l}, \varphi_{k m}\}=\delta_{jk}\delta_{lm}.
\end{eqnarray*}

Let us here consider a specific series of operators 
\begin{eqnarray}
\eta_{2j-1\: l}
=
\sigma^{z}_{j\: l}, 
\hspace{0.6cm}
\eta_{2j\: l}
=
\sigma^{x}_{j\: l}\sigma^{x}_{j+1\: l}, 
\label{seriesTrIsing}
\end{eqnarray}
together with the initial operators $\eta_{0l}=i\sigma^{x}_{1 l}$.  
The index $j$ runs $1\leq j\leq N$, 
where $N$ is the number of sites in a row, 
and in this case we have $M=2N$. 
Then the transformations, for example with $l=2$ and $1$, 
are schematically written as 
\begin{eqnarray}
\varphi_{2j-1\: 2}
&=&
\frac{1}{\sqrt{2}}\times 
\begin{array}{l}
\sigma^{z}_{12}\sigma^{z}_{22}\cdots \sigma^{z}_{j-1\:2}\sigma^{y}_{j2}\\
\sigma^{z}_{11}\sigma^{z}_{21}\cdots \sigma^{z}_{j-1\:1}\sigma^{z}_{j1}\cdots \sigma^{z}_{k1}\cdots \sigma^{z}_{N1},
\end{array}
\nonumber\\
\varphi_{2j-2\: 2}
&=&
\frac{1}{\sqrt{2}}\times 
\begin{array}{l}
\sigma^{z}_{12}\sigma^{z}_{22}\cdots \sigma^{z}_{j-1\:2}\sigma^{x}_{j2}\\
\sigma^{z}_{11}\sigma^{z}_{21}\cdots \sigma^{z}_{j-1\:1}\sigma^{z}_{j1}\cdots \sigma^{z}_{k1}\cdots \sigma^{z}_{N1},
\end{array}
\nonumber\\
\varphi_{2k-2\: 1}
&=&
\frac{1}{\sqrt{2}}\times 
\begin{array}{l}
\hspace{0.6cm}
 \\
\sigma^{z}_{11}\sigma^{z}_{21}\cdots \sigma^{z}_{j-1\:1}\sigma^{z}_{j1}\cdots \sigma^{x}_{k1}.
\end{array}
\label{JordanWignerExample}
\end{eqnarray}
One may readily verify that the anti-commutation relation 
$\{\varphi_{j l}, \varphi_{k m}\}=\delta_{jk}\delta_{lm}$ 
comes from the anti-commutation relations 
$\{\sigma^{y}_{j2}, \sigma^{x}_{j2}\}=0$ and 
$\{\sigma^{z}_{k1}, \sigma^{x}_{k1}\}=0$. 
This is the Jordan-Wigner transformation in two-dimension
\cite{07Feng}-\cite{08ChenNussinov}, 
i.e. the two-dimensional Jordan-Wigner transformation is obtained 
as a special case of (\ref{transgenrow}).


\section{Operators and interactions}\label{OperatorsAndInteractions}

\begin{figure}
\includegraphics{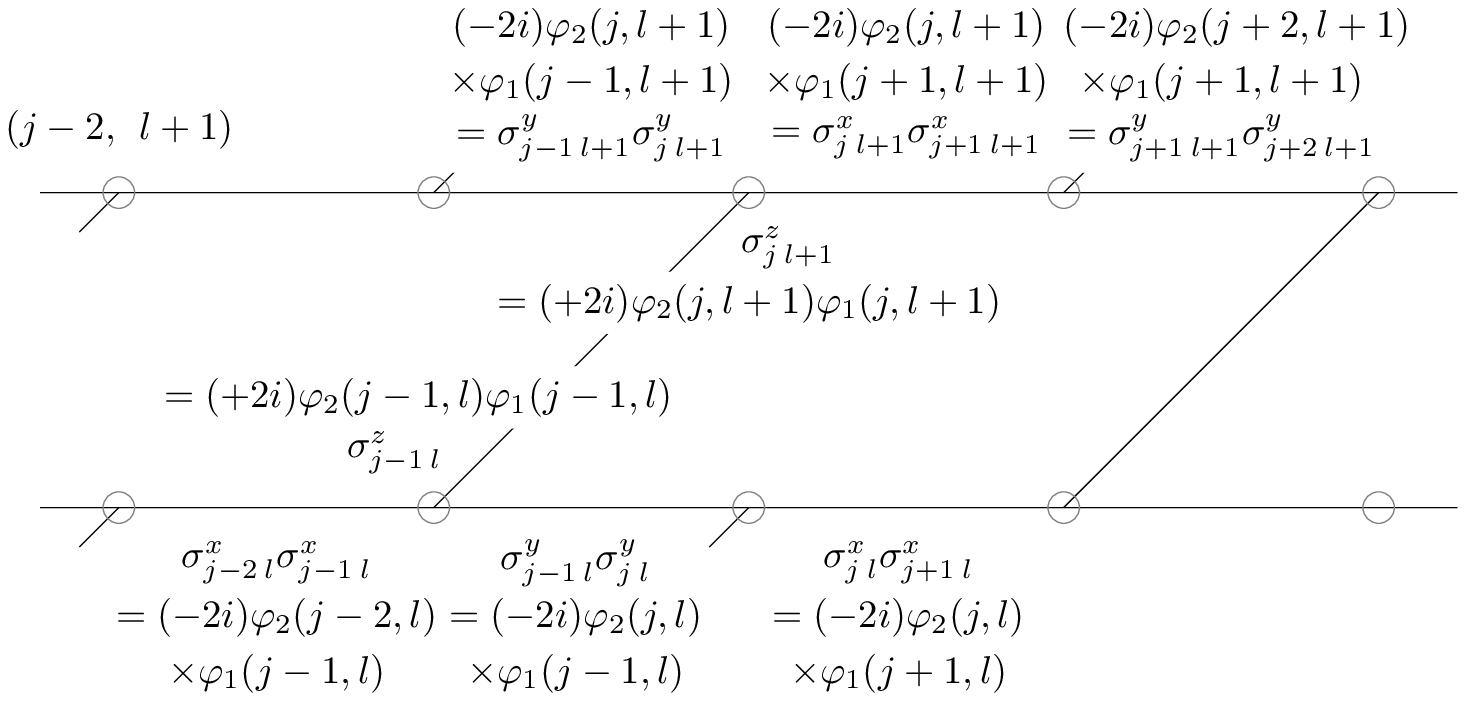}
\caption{\label{fig:Operators}Spin-spin interactions and corresponding Majorana operators $\varphi_{\alpha}(j, l)$ on the lattice. }
\end{figure}

When we consider the series of operators (\ref{seriesTrIsing}),  
we find from (\ref{JordanWignerExample}), that 
$\varphi_{\rho l}$ are classified into two kinds of operators 
according to $\rho=$odd and $\rho=$even. 
Thus we will introduce a new notation 
\begin{eqnarray}
\varphi_{1}(j, l)=\varphi_{2j-2\: l}, 
\hspace{0.6cm}
\varphi_{2}(j, l)=\varphi_{2j-1\: l}. 
\label{newnotation}
\end{eqnarray}
The operators and their relations are summarized 
in Table 1 and Fig.\ref{fig:Operators}. 
Interactions in (\ref{HamiltonianKitaev}) are expressed, 
in terms of $\varphi_{\alpha}(j, l)$, as 
\begin{eqnarray}
K_{x}\sigma^{x}_{j\: l}\sigma^{x}_{j+1\: l}
&=&
K_{x}(-2i)\:\varphi_{2}(j, l)\varphi_{1}(j+1, l),
\nonumber\\
K_{y}\sigma^{y}_{j-1\: l}\sigma^{y}_{j\: l}
&=&
K_{y}(-2i)\:\varphi_{2}(j, l)\varphi_{1}(j-1,\: l),
\nonumber\\
K_{z}\sigma^{z}_{j-1\: l}\sigma^{z}_{j \:l+1}
&=&
K_{z}
(+2i)\:\varphi_{2}(j-1, l)\varphi_{1}(j-1,\: l)
\nonumber\\
&&
\times (+2i)\:\varphi_{2}(j, l+1)\varphi_{1}(j,\: l+1).
\label{KxKyKzvarphi}
\end{eqnarray}
The first two interactions consist of 
two-body products of the operators $\varphi_{\alpha}(j, l)$ 
with uniform coupling constants, 
and thus can be diagonalized exactly.

The interaction $K_{z}\sigma^{z}_{j-1\: l}\sigma^{z}_{j\: l+1}$ 
is expressed as a four-body product of $\varphi_{\alpha}(j, l)$. 
We can find, however, from Fig.\ref{fig:Operators} 
that $\varphi_{2}(j-1, l)\varphi_{1}(j, l+1)$ 
is disjoint from other operators, 
and takes one of its eigenvalues, like a 'floating spin'. 
We find 
$(\varphi_{2}(j-1, l)\varphi_{1}(j, l+1))^{2}
=
-\varphi_{2}(j-1, l)^{2}\varphi_{1}(j, l+1)^{2}
=
-(1/2)^{2}$, 
and then the eigenvalues are $\pm i/2$.  
Hence the product of operators $\varphi_{2}(j-1, l)\varphi_{1}(j, l+1)$ 
works as a constant in each eigenspace, 
and $K_{z}\sigma^{z}_{j-1\: l}\sigma^{z}_{j l+1}$ 
is expressed as 
\begin{eqnarray*}
K_{z}\sigma^{z}_{j-1\: l}\sigma^{z}_{j\: l+1}
=
K_{z}
(+4\Psi_{jl})\:\varphi_{2}(j, l+1)\varphi_{1}(j-1, l),
\end{eqnarray*}
where $\Psi_{jl}=\varphi_{2}(j-1, l)\varphi_{1}(j, l+1)$.

The operators $\Psi_{jl}$ commute with all the interactions 
in (\ref{HamiltonianKitaev}), (\ref{Hamiltonian3}), and (\ref{HamiltonianWen}),
and have simple relation with the operators $W_{jl}$ given in (\ref{Wjl}). 
Let us express $\Psi_{jl}$ in terms of the spin operators as 
\begin{eqnarray}
\Psi_{jl}
&=&
\varphi_{2}(j-1, l)\varphi_{1}(j, l+1)
\nonumber\\
&=&
\varphi_{2j-3\: l}\varphi_{2j-2\: l+1}
\nonumber\\
&=&
\frac{i}{2}\times 
\begin{array}{l}
\sigma^{z}_{1\:l+1}\sigma^{z}_{2\:l+1}\cdots \sigma^{z}_{j-1\:l+1}\sigma^{x}_{j\:l+1}\\
\hspace{2.5cm}
\sigma^{x}_{j-1\:l}
\hspace{0.4cm}
\sigma^{z}_{j\:l}\cdots \sigma^{z}_{N\:l}.
\end{array}
\label{PsijlSpin}
\end{eqnarray}
From (\ref{PsijlSpin}), we find 
\begin{eqnarray*}
\Psi_{jl}\Psi_{j+2\: l}
&=&
\varphi_{2j-3\: l}\varphi_{2j-2\: l+1}\cdot\varphi_{2j+1\: l}\varphi_{2j+2\: l+1}
\\
&=&
-\frac{1}{4}
\sigma^{x}_{j-1\:l}\sigma^{z}_{j\:l}(\sigma^{z}_{j+1\:l}\sigma^{x}_{j+1\:l})
(\sigma^{x}_{j\:l+1}\sigma^{z}_{j\:l+1})\sigma^{z}_{j+1\:l+1}\sigma^{x}_{j+2\:l+1}
\\
&=&
-\frac{1}{4}
\sigma^{x}_{j-1\:l}\sigma^{z}_{j\:l}\sigma^{y}_{j+1\:l}
\sigma^{y}_{j\:l+1}\sigma^{z}_{j+1\:l+1}\sigma^{x}_{j+2\:l+1}
\\
&=&
-\frac{1}{4}W_{jl}.
\end{eqnarray*}

\begin{figure}
\includegraphics{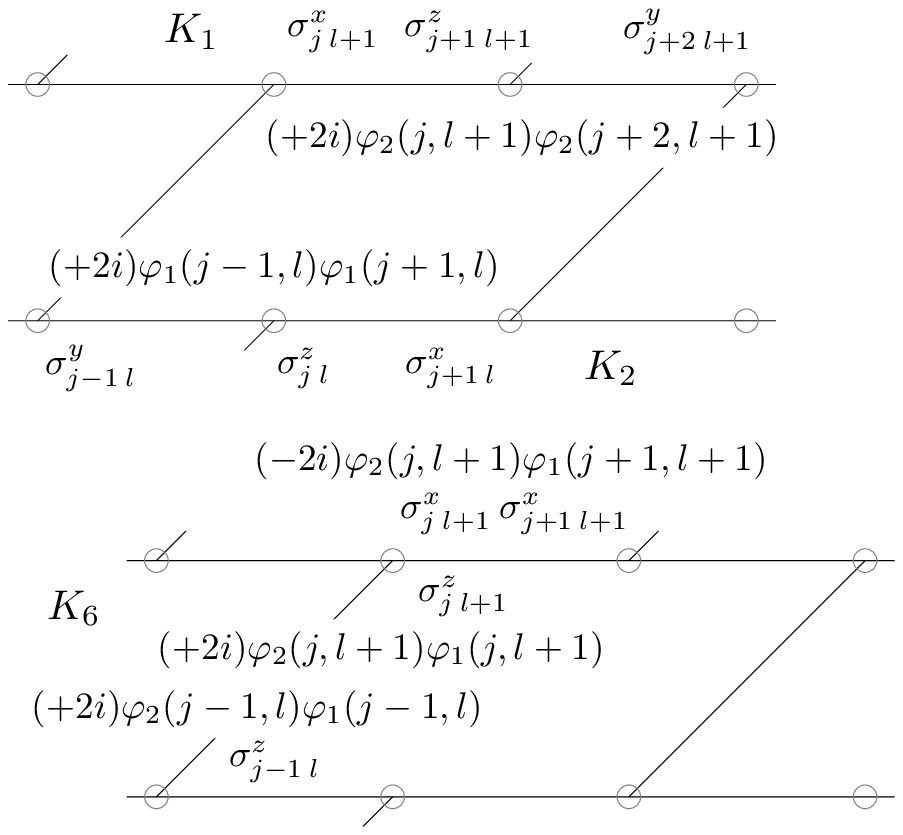}
\caption{\label{fig:3bodyInteractions}Three-body interactions $K_{1}$, $K_{2}$, and $K_{6}$.}
\end{figure}

\begin{figure}
\includegraphics{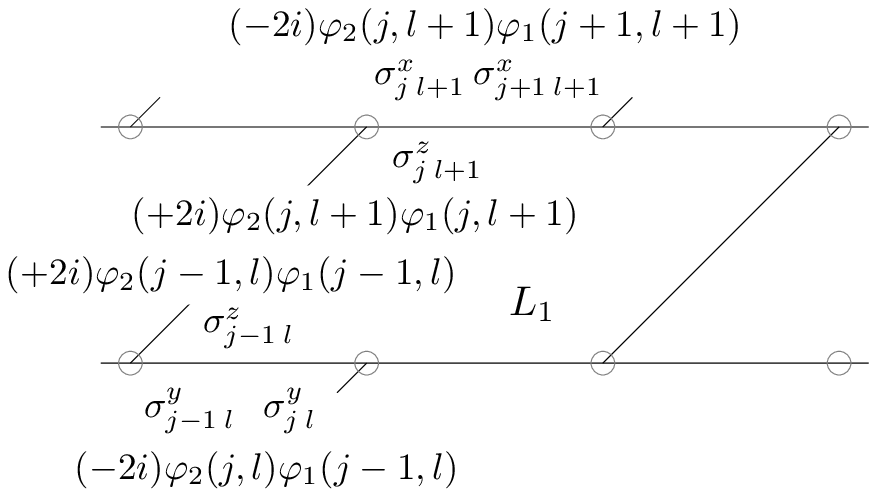}
\caption{\label{fig:4bodyInteractions}Four-body interaction $L_{1}$.}
\end{figure}

Let us next consider the three-body interactions 
shown in Fig.\ref{fig:3bodyInteractions}. 
The interactions are expressed in terms of $\varphi_{\alpha}(j, l)$ as
\begin{eqnarray*}
K_{1}\sigma^{x}_{j\: l}\sigma^{z}_{j+1\: l}\sigma^{y}_{j+2\: l}
&=&
(-2i)\:\varphi_{2}(j, l)\varphi_{2}(j+2, l),
\nonumber\\
K_{2}\sigma^{y}_{j-1\: l}\sigma^{z}_{j l}\sigma^{x}_{j+1\: l}
&=&
(+2i)\:\varphi_{1}(j-1, l)\varphi_{1}(j+1, l),
\end{eqnarray*}
and, for example,   
\begin{eqnarray}
K_{6}
\sigma^{z}_{j-1\: l}\sigma^{y}_{j\: l+1}\sigma^{x}_{j+1\: l+1}
&=&
K_{6}
(-i)\sigma^{z}_{j-1\: l}
\cdot \sigma^{z}_{j\: l+1}
\cdot \sigma^{x}_{j\: l+1}\sigma^{x}_{j+1\: l+1}
\nonumber\\
&=&
K_{6}(-i)
(+2i)\:\varphi_{2}(j-1, l)\varphi_{1}(j-1, l)
\nonumber\\
&&
\hspace{1.0cm}
\times
(+2i)\:\varphi_{2}(j, l+1)\varphi_{1}(j, l+1)
\nonumber\\
&&
\hspace{1.0cm}
\times
(-2i)\:\varphi_{2}(j, l+1)\varphi_{1}(j+1, l+1).
\label{L3int}
\end{eqnarray}
We again find 
$\varphi_{2}(j-1, l)\varphi_{1}(j, l+1)=\Psi_{jl}$, 
which works as a constant, 
and also 
$\varphi_{2}(j, l+1)^{2}=1/2$. 
Then (\ref{L3int}) can be expressed as 
\begin{eqnarray*}
&&
K_{6}
\sigma^{z}_{j-1\: l}\sigma^{y}_{j\: l+1}\sigma^{x}_{j+1\: l+1}
=
K_{6}
(+4\Psi_{jl})
\varphi_{1}(j-1, l)\varphi_{1}(j+1, l+1).
\end{eqnarray*}
One can similarly introduce other three-body interactions as follows: 
\begin{eqnarray*}
K_{3}
\sigma^{y}_{j\: l}\sigma^{x}_{j-1\: l}\sigma^{z}_{j\: l+1}
&=&
K_{3}(-i)
\sigma^{y}_{j-1\: l}\sigma^{y}_{j\: l}
\cdot \sigma^{z}_{j-1\: l}\sigma^{z}_{j\: l+1}
\\
&=&
K_{3}(+4\Psi_{jl})
\varphi_{2}(j, l)\varphi_{2}(j, l+1),
\end{eqnarray*}
\begin{eqnarray*}
K_{4}
\sigma^{z}_{j-1\: l}\sigma^{x}_{j\: l+1}\sigma^{y}_{j-1\: l+1}
&=&
K_{4}(+i)
\sigma^{z}_{j-1\: l}\sigma^{z}_{j\: l+1}
\cdot \sigma^{y}_{j-1\: l+1}\sigma^{y}_{j\: l+1}
\\
&=&
K_{4}(-4\Psi_{jl})
\varphi_{1}(j-1, l)\varphi_{1}(j-1, l+1),
\nonumber
\end{eqnarray*}
and
\begin{eqnarray*}
K_{5}
\sigma^{x}_{j-2\: l}\sigma^{y}_{j-1\: l}\sigma^{z}_{j\: l+1}
&=&
K_{5}
(+i)\sigma^{x}_{j-2\: l}\sigma^{x}_{j-1\: l}
\cdot \sigma^{z}_{j-1\: l}\cdot\sigma^{z}_{j\: l+1}
\nonumber\\
&=&
K_{5}
(-4\Psi_{jl})
\varphi_{2}(j-2, l)\varphi_{2}(j, l+1).
\nonumber
\end{eqnarray*}
The  four-body interaction $L_{1}$
is expressed, as shown in Fig.\ref{fig:4bodyInteractions}, as  
\begin{eqnarray*}
L_{1}
\sigma^{y}_{j\: l+1}\sigma^{x}_{j+1\: l+1}
\sigma^{x}_{j-1\: l}\sigma^{y}_{j\: l}
&=&
L_{1}
\sigma^{z}_{j\: l+1}
\cdot \sigma^{x}_{j\: l+1}\sigma^{x}_{j+1\: l+1}
\cdot \sigma^{z}_{j-1\: l}
\cdot \sigma^{y}_{j-1\: l}\sigma^{y}_{j\: l}
\nonumber\\
&=&
L_{1}
(+2i)\:\varphi_{2}(j, l+1)\varphi_{1}(j, l+1)
\nonumber\\
&&
\hspace{0.1cm}
\times
(-2i)\:\varphi_{2}(j, l+1)\varphi_{1}(j+1, l+1)
\nonumber\\
&&
\hspace{0.1cm}
\times
(+2i)\:\varphi_{2}(j-1, l)\varphi_{1}(j-1, l)
\nonumber\\
&&
\hspace{0.1cm}
\times
(-2i)\:\varphi_{2}(j, l)\varphi_{1}(j-1, l)
\nonumber\\
&=&
L_{1}(-4\Psi_{jl})
\varphi_{2}(j, l)\varphi_{1}(j+1, l+1),
\nonumber
\end{eqnarray*}
and another interaction $L_{2}$ is 
\begin{eqnarray*}
L_{2}
\sigma^{y}_{j+1\: l+1}\sigma^{x}_{j+2\: l+1}
\sigma^{x}_{j\: l}\sigma^{y}_{j+1\: l}
&=&
L_{2}
\sigma^{y}_{j+1\: l+1}\sigma^{y}_{j+2\: l+1}
\cdot \sigma^{z}_{j+2\: l+1}
\cdot \sigma^{x}_{j\: l}\sigma^{x}_{j+1\: l}
\cdot \sigma^{z}_{j+1\: l}
\nonumber\\
&=&
L_{2}(-4\Psi_{j+2\:l})
\varphi_{2}(j, l)\varphi_{1}(j+1, l+1).
\nonumber
\end{eqnarray*}

\section{Generalizations}\label{Generalizations}

We introduced a series of operators 
$\eta_{2j-1}=(+2i)\varphi_{2}(j)\varphi_{1}(j)$ and 
$\eta_{2j}=(-2i)\varphi_{2}(j)\varphi_{1}(j+1)$. 
Let us then consider another series of operators  
\begin{eqnarray*}
\eta_{1},
\hspace{0.3cm}
\eta_{2}\eta_{3}\eta_{4},
\hspace{0.3cm}
\eta_{5},
\hspace{0.3cm}
\eta_{6}\eta_{7}\eta_{8},
\hspace{0.3cm}
\eta_{9},
\hspace{0.3cm}
\ldots, 
\end{eqnarray*}
and generally
\begin{eqnarray}
{\bar \eta}_{2j-1}
&=&
\eta_{4j-3}
\hspace{1.4cm}
=
(+2i)\varphi_{2}(2j-1)\varphi_{1}(2j-1),
\nonumber\\
{\bar \eta}_{2j}
&=&
\eta_{4j-2}\eta_{4j-1}\eta_{4j}
=
(-2i)\varphi_{2}(2j-1)\varphi_{1}(2j+1).
\label{baretajex}
\end{eqnarray}
The series $\{{\bar \eta}_{j}\}$ satisfies the condition (\ref{cond}) 
and generates solvable Hamiltonians. 
Similarly let us consider series of operators 
with periodic structures as 
\begin{eqnarray}
&&
\varphi_{2}(\rho)\varphi_{1}(\rho+k),  
\hspace{1.9cm}
\varphi_{2}(\rho)\varphi_{1}(\rho+k+l), 
\nonumber\\
&&
\varphi_{2}(\rho+l)\varphi_{1}(\rho+k+l), 
\hspace{0.6cm}
\varphi_{2}(\rho+l)\varphi_{1}(\rho+k+2l), 
\nonumber\\
&&
\varphi_{2}(\rho+2l)\varphi_{1}(\rho+k+2l), 
\hspace{0.2cm}
\varphi_{2}(\rho+2l)\varphi_{1}(\rho+k+3l), 
\nonumber\\
&&
\hspace{0.8cm}
\ldots
\label{seriesperiodic}
\end{eqnarray}
Generally, we can introduce series of operators 
that satisfy (\ref{cond}) as
\begin{eqnarray}
{\bar \eta}_{2j-1}
&=&
(\pm 2i)
\varphi_{2}(\rho+(j-1)l)\varphi_{1}(\rho+(j-1)l+k), 
\nonumber\\
{\bar \eta}_{2j}
&=&
(\pm 2i)
\varphi_{2}(\rho+(j-1)l)\varphi_{1}(\rho+jl+k), 
\label{seriesperiodicgen}
\end{eqnarray}
where the sign of the factors $(\pm 2i)$ are arbitrary, 
$l$ and $k$ are integers, $l\geq 1$, 
and $\rho=1, 2, \ldots, l$ 
is fixed in each series. 
The series with different $\rho$ commute with each other. 

More generally we can consider, from (\ref{transmain}), series of operators  
\begin{eqnarray}
\varphi_{\tau_{1}}\varphi_{\tau_{2}},
\hspace{0.3cm}
\varphi_{\tau_{2}}\varphi_{\tau_{3}},
\hspace{0.3cm}
\varphi_{\tau_{3}}\varphi_{\tau_{4}},
\hspace{0.3cm}
\varphi_{\tau_{4}}\varphi_{\tau_{5}},
\hspace{0.3cm}
\ldots
\label{seriesgen}
\end{eqnarray}
and generally introduce 
${\bar \eta}_{j}=(\pm 2i)\varphi_{\tau_{j}}\varphi_{\tau_{j+1}}$, 
where all $\tau_{j}$ are different with each other. 
Then the series (\ref{seriesgen}) satisfies (\ref{cond}).

\section{Equivalent Hamiltonian}\label{EquivalentHamiltonian}

As an example of (\ref{seriesperiodicgen}), 
let us consider (\ref{baretajex}), 
which is a special case of (\ref{seriesperiodicgen}) 
with $\rho=1$, $l=2$, and $k=0$. 
Let us consider the series of operators (\ref{seriesTrIsing}) in Table 1, 
and in this case we have 
\begin{eqnarray*}
{\bar \eta}_{1l}=\sigma^{z}_{1l}, 
\hspace{0.3cm}
{\bar \eta}_{2l}=-\sigma^{x}_{1l}\sigma^{z}_{2l}\sigma^{x}_{3l},
\hspace{0.3cm}
{\bar \eta}_{3l}=\sigma^{z}_{3l}, 
\hspace{0.3cm}
{\bar \eta}_{4l}=-\sigma^{x}_{3l}\sigma^{z}_{4l}\sigma^{x}_{5l}, 
\nonumber\\
\end{eqnarray*}
and generally 
\begin{eqnarray*}
{\bar \eta}_{2j-1\: l}=\sigma^{z}_{2j-1\: l},
\hspace{0.2cm}
{\bar \eta}_{2j\: l}=-\sigma^{x}_{2j-1\: l}\:\sigma^{z}_{2j\: l}\:\sigma^{x}_{2j+1\: l}.
\end{eqnarray*}
The initial operators can be chosen as ${\bar \eta}_{0l}=i\sigma^{x}_{1l}$. 
Following (\ref{HamiltonianKitaev}) and (\ref{KxKyKzvarphi}),
we will introduce the Hamiltonians 
\begin{eqnarray*}
{\cal H}_{x}
&=&
K_{x}\sum_{l=1}^{M_{2}}\sum_{j=odd}^{M_{1}-1}
(-2i){\bar \varphi}_{2}(j, l){\bar \varphi}_{1}(j+1, l)
\nonumber\\
&=&
K_{x}\sum_{l=1}^{M_{2}}\sum_{j=odd}^{M_{1}-1}
(-\sigma^{x}_{2j-1\: l}
\sigma^{z}_{2j\: l}
\sigma^{x}_{2j+1\: l}),
\nonumber\\
{\cal H}_{y}
&=&
K_{y}\sum_{l=1}^{M_{2}}\sum_{j=odd}^{M_{1}-1}
(-2i){\bar \varphi}_{2}(j, l){\bar \varphi}_{1}(j-1, l)
\nonumber\\
&=&
K_{y}\sum_{l=1}^{M_{2}}\sum_{j=odd}^{M_{1}-1}
(-\sigma^{y}_{2j-3\: l}
\sigma^{z}_{2j-2\: l}
\sigma^{y}_{2j-1\: l}),
\nonumber\\
{\cal H}_{z}
&=&
K_{z}\sum_{l=1}^{M_{2}}\sum_{j=odd}^{M_{1}-1}
(+2i){\bar \varphi}_{2}(j-1, l){\bar \varphi}_{1}(j-1, l)
\nonumber\\
&&
\hspace{1.6cm}
\times
(+2i){\bar \varphi}_{2}(j, l+1){\bar \varphi}_{1}(j, l+1)
\nonumber\\
&=&
K_{z}\sum_{l=1}^{M_{2}}\sum_{j=odd}^{M_{1}-1}
\sigma^{z}_{2j-3\: l}
\sigma^{z}_{2j-1\: l+1},
\end{eqnarray*}
where ${\bar \varphi}_{\alpha}(j, l)$ 
are obtained from (\ref{transmain}) and (\ref{newnotation})
replacing $\eta_{jl}$ by ${\bar \eta}_{jl}$. 

\begin{figure}
\includegraphics{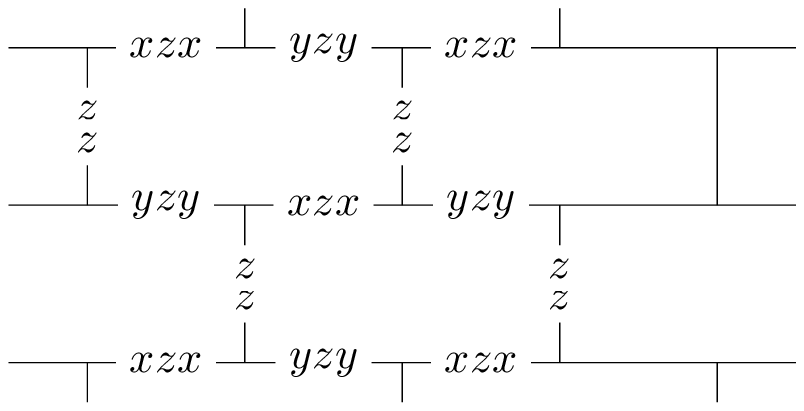}
\caption{\label{fig:EquivHamiltonian} A model equivalent to the honeycomb lattice Kitaev model. 
The model is composed of the cluster and the Ising interactions. }
\end{figure}

The interactions are shown in Fig.\ref{fig:EquivHamiltonian}. 
The sum ${\cal H}_{x}+{\cal H}_{y}$ is the Hamiltonian of parallel spin chains 
with the cluster-type interactions, 
and ${\cal H}_{z}$ is the Hamiltonian of the Ising interactions 
between these chains.  
The total Hamiltonian ${\cal H}_{K2}={\cal H}_{x}+{\cal H}_{y}+{\cal H}_{z}$ 
is equivalent to the honeycomb lattice Kitaev model (\ref{HamiltonianKitaev}) 
(there is no difference 
when two Hamiltonians are written in terms of 
$\varphi_{\alpha}(j, l)$ and ${\bar \varphi}_{\alpha}(j, l)$). 
We can also find other equivalent Hamiltonians 
from the series of operators (\ref{seriesClusterNnn}) in Table 1.

\section{Diagonalization}\label{Diagonalization}

We will derive the phase structure of the case 
with the interactions $K_{x}$, $K_{y}$, $K_{z}$, and $L_{1}$ and $L_{2}$. 
In this case, Lieb's theorem\cite{94Lieb} applies 
and it is proved that one of the ground states is found in the subspace 
where $\Psi_{jl}=-i/2$ for all $j$ and $l$. 
In this subspace, the translational invariance of $\Psi_{jl}$ 
enables us to derive the ground state energy explicitly. 

We need $\varphi_{2}(j, l)$ with odd $j$, and $\varphi_{1}(j, l)$ with even $j$. 
Let us consider Fourier transformations 
\begin{eqnarray}
\varphi_{2}(j, l)
=
\frac{1}{\sqrt{\frac{M_{1}}{2}}}\frac{1}{\sqrt{M_{2}}}
\sum_{
\stackrel
{\mbox{$\scriptstyle -\pi\leq q_{1}<\pi$}}
{\mbox{$\scriptstyle -\pi\leq q_{2}<\pi$}}
}
e^{iq_{1}k}e^{iq_{2}l}c_{2}(q_{1}, q_{2}), 
\label{FourierTranformation2}
\end{eqnarray}
where $j=2k-1$ is odd, and 
\begin{eqnarray}
\varphi_{1}(j, l)
=
\frac{1}{\sqrt{\frac{M_{1}}{2}}}\frac{1}{\sqrt{M_{2}}}
\sum_{
\stackrel
{\mbox{$\scriptstyle -\pi\leq q_{1}<\pi$}}
{\mbox{$\scriptstyle -\pi\leq q_{2}<\pi$}}
}
e^{iq_{1}k}e^{iq_{2}l}e^{iq_{1}/2}c_{1}(q_{1}, q_{2}), 
\label{FourierTranformation1}
\end{eqnarray}
where $j=2k$ is even. 
The factor $e^{iq_{1}/2}$ in (\ref{FourierTranformation1}) 
is introduced so as to have a symmetric form of the Hamiltonian 
later in (\ref{hammatrix}). 
The operators $c_{\alpha}(q_{1}, q_{2})$ 
are the fermi operators satisfying 
$\{c^{\dag}_{\alpha}(p_{1}, p_{2}), c_{\beta}(q_{1}, q_{2})\}
=\delta_{\alpha\beta}\delta_{p_{1}q_{1}}\delta_{p_{2}q_{2}}$  
and 
$\{c_{\alpha}(p_{1}, p_{2}), c_{\beta}(q_{1}, q_{2})\}=0$.  
The inverse transformation for $\varphi_{2}(j, l)$ is  
\begin{eqnarray*}
c_{2}(q_{1}, q_{2})
=
\frac{1}{\sqrt{\frac{M_{1}}{2}}}\frac{1}{\sqrt{M_{2}}}
\sum_{k=1}^{M_{1}/2}
\sum_{l=1}^{M_{2}}
e^{-iq_{1}k}e^{-iq_{2}l}\varphi_{2}(j, l).
\end{eqnarray*}
A similar relation appears for $\varphi_{1}(j, l)$ 
with an additional phase factor $e^{-iq_{1}/2}$. 
The operators $\varphi_{\alpha}(j, l)$ $(\alpha=1, 2)$ 
defined in (\ref{transgenrow}), (\ref{seriesTrIsing}) and (\ref{newnotation})
satisfy 
$\varphi_{\alpha}^{\dag}(j, l)=\varphi_{\alpha}(j, l)$, 
and thus we find that $c_{\alpha}(-q_{1}, -q_{2})=c^{\dag}_{\alpha}(q_{1}, q_{2})$. 
Then the term proportional to $K_{x}$ in (\ref{HamiltonianKitaev}) is expressed, 
from (\ref{KxKyKzvarphi}), (\ref{FourierTranformation2}) and (\ref{FourierTranformation1}), as 
\begin{eqnarray}
&&
K_{x}(-2i)
\sum_{
\stackrel
{\mbox{$\scriptstyle -\pi\leq q_{1}<\pi$}}
{\mbox{$\scriptstyle -\pi\leq q_{2}<\pi$}}
}
c_{2}(q_{1},q_{2})e^{-iq_{1}/2}c_{1}^{\dag}(q_{1},q_{2}).
\nonumber\\
&=&
K_{x}(-2i)\:\:
\frac{1}{2}\sum_{
\stackrel
{\mbox{$\scriptstyle -\pi\leq q_{1}<\pi$}}
{\mbox{$\scriptstyle -\pi\leq q_{2}<\pi$}}
}
(
c_{2}(q_{1},q_{2})e^{-iq_{1}/2}c^{\dag}_{1}(q_{1},q_{2})
+
c^{\dag}_{2}(q_{1},q_{2})e^{iq_{1}/2}c_{1}(q_{1},q_{2})
).
\nonumber\\
\label{KxSym}
\end{eqnarray}
Note that 
$c_{\alpha}(-q_{1}, -q_{2})\:(=c^{\dag}_{\alpha}(q_{1}, q_{2}))$ 
and $c_{\alpha}(q_{1}, q_{2})$ 
creates and annihilates the same particle 
indexed by $\alpha$ and $(q_{1}, q_{2})$. 
Thus the summation should be restricted to the region, for example,  
$0\leq q_{1}<\pi$ and $-\pi\leq q_{2}<\pi$, 
instead of $-\pi\leq q_{1}<\pi$ and $-\pi\leq q_{2}<\pi$. 
The term (\ref{KxSym}) is then written as 
\begin{eqnarray}
K_{x}(-2i)
\sum_{
\stackrel
{\mbox{$\scriptstyle 0\leq q_{1}<\pi$}}
{\mbox{$\scriptstyle -\pi\leq q_{2}<\pi$}}
}
(
c_{2}(q_{1},q_{2})e^{-iq_{1}/2}c^{\dag}_{1}(q_{1},q_{2})
+
c^{\dag}_{2}(q_{1},q_{2})e^{iq_{1}/2}c_{1}(q_{1},q_{2})
).
\label{Kx2}
\end{eqnarray}
The total Hamiltonian is expressed as 
\begin{eqnarray*}
&&
-\beta{\cal H}
=
\sum_{
\stackrel
{\mbox{$\scriptstyle 0\leq q_{1}<\pi$}}
{\mbox{$\scriptstyle -\pi\leq q_{2}<\pi$}}
}
\Big[
h_{11}c^{\dag}_{1}(q_{1},q_{2})c_{1}(q_{1},q_{2})
+
h_{22}c^{\dag}_{2}(q_{1},q_{2})c_{2}(q_{1},q_{2})
\\
&&
\hspace{2.8cm}
+
h_{12}c^{\dag}_{1}(q_{1},q_{2})c_{2}(q_{1},q_{2})
+
h_{21}c^{\dag}_{2}(q_{1},q_{2})c_{1}(q_{1},q_{2})
\Big],
\end{eqnarray*}
where 
\begin{eqnarray}
h_{11}
&=&
K_{2}(-2i)(e^{iq_{1}}-e^{-iq_{1}})
+
K_{4}(-4\Psi)(e^{iq_{2}}-e^{-iq_{2}})
+
K_{6}(+4\Psi)(e^{iq_{1}+iq_{2}}-e^{-iq_{1}-iq_{2}}),
\nonumber\\
h_{22}
&=&
K_{1}(-2i)(e^{iq_{1}}-e^{-iq_{1}})
+
K_{3}(+4\Psi)(e^{iq_{2}}-e^{-iq_{2}})
+
K_{5}(-4\Psi)(e^{iq_{1}+iq_{2}}-e^{-iq_{1}-iq_{2}}),
\nonumber\\
h_{21}
&=&
+K_{x}(-2i)e^{iq_{1}/2}
+K_{y}(-2i)e^{-iq_{1}/2}
+
K_{z}(+4\Psi)e^{-iq_{1}/2}e^{-iq_{2}}
+
(L_{1}+L_{2})(-4\Psi)e^{iq_{1}/2}e^{iq_{2}},
\nonumber\\
h_{12}
&=&
h^{\dag}_{21},
\label{hammatrixelements}
\end{eqnarray}
and $\Psi_{jl}=\Psi=-i/2$. 
The Hamiltonian is also expressed, 
with the basis states 
$c^{\dag}_{2}(q_{1},q_{2})c^{\dag}_{1}(q_{1},q_{2})|0\rangle$, 
$c^{\dag}_{2}(q_{1},q_{2})|0\rangle$, 
$c^{\dag}_{1}(q_{1},q_{2})|0\rangle$, 
and $|0\rangle$, 
where $|0\rangle$ is the vacuum, as 
\begin{eqnarray}
-\beta{\cal H}
=
\sum_{
\stackrel
{\mbox{$\scriptstyle 0\leq q_{1}<\pi$}}
{\mbox{$\scriptstyle -\pi\leq q_{2}<\pi$}}
}
\left(
\begin{array}{cccc}
h_{11}+h_{22}&0&0&0\\
0&h_{22}&h_{21}&0\\
0&h_{12}&h_{11}&0\\
0&0&0&0
\end{array}
\right).
\label{hammatrix}
\end{eqnarray}
In our case, where the three-body terms are absent, 
the energy eigenvalues are 
\begin{eqnarray}
\lambda
=
0, 
\hspace{0.2cm}
0,
\hspace{0.2cm}
{\rm and} 
\hspace{0.2cm}
\pm\sqrt{4kk^{*}},
\label{eigenvalues}
\end{eqnarray}
where
\begin{eqnarray}
k
&=&
K_{x}e^{iq_{1}/2}+K_{y}e^{-iq_{1}/2}+K_{z}e^{-i(q_{1}/2+q_{2})}
+Le^{i(q_{1}/2+q_{2})},
\nonumber\\
L
&=&
-(L_{1}+L_{2}).
\label{eigenvaluesk}
\end{eqnarray}

\section{Gapless condition}\label{GaplessCondition}

We will consider the gapless condition 
that $\sqrt{4kk^{*}}$ in (\ref{eigenvalues}) becomes zero with some $q_{1}$ and $q_{2}$.  
Let $K_{x}=\beta J_{x}$, $K_{y}=\beta J_{y}$, $K_{z}=\beta J_{z}$, 
and $L=\beta J_{4}$. 
Four terms from (\ref{eigenvaluesk}) ,
\begin{eqnarray*}
J_{x}e^{iq_{1}/2}+J_{y}e^{-iq_{1}/2}
\hspace{0.3cm}
{\rm and} 
\hspace{0.3cm}
J_{z}e^{-i(q_{1}/2+q_{2})}+J_{4}e^{i(q_{1}/2+q_{2})},
\end{eqnarray*}
form two ellipses on the complex plane of the variable $z=x+iy$.  
The first two terms including $J_{x}$ and $J_{y}$ form 
\begin{eqnarray}
\frac{x^{2}}{(J_{x}+J_{y})^{2}}
+\frac{y^{2}}{(J_{x}-J_{y})^{2}}
=1,
\label{ellipse1}
\end{eqnarray}
where $0\leq q_{1}/2<\pi/2$ corresponds to a part of the ellipse. 
The latter two terms including $J_{z}$ and $J_{4}$ form 
\begin{eqnarray}
\frac{x^{2}}{(J_{4}+J_{z})^{2}}
+\frac{y^{2}}{(J_{4}-J_{z})^{2}}
=1, 
\label{ellipse2}
\end{eqnarray}
where $0\leq q_{1}/2<\pi/2$ and $-\pi\leq q_{2}<\pi$ 
corresponds to the full ellipse. 
The condition is satisfied 
if (\ref{ellipse1}) and (\ref{ellipse2}) 
are simultaneously satisfied with some real $(x, y)$. 
From (\ref{ellipse1}) and (\ref{ellipse2}) 
we obtain 
\begin{eqnarray}
\Phi(-+)\: x^{2}
=
\phi(-),
\hspace{0.4cm}
\Phi(+-)\: y^{2}
=
\phi(+),
\label{x2y2cond}
\end{eqnarray}
where
\begin{eqnarray*}
\Phi(-+)
&=&
\Big(\frac{J_{x}-J_{y}}{J_{x}+J_{y}}\Big)^{2}
-
\Big(\frac{J_{4}-J_{z}}{J_{4}+J_{z}}\Big)^{2},
\\
\Phi(+-)
&=&
\Big(\frac{J_{x}+J_{y}}{J_{x}-J_{y}}\Big)^{2}
-
\Big(\frac{J_{4}+J_{z}}{J_{4}-J_{z}}\Big)^{2},
\\
\phi(\pm)
&=&
(J_{x}\pm J_{y})^{2}-(J_{4}\pm J_{z})^{2}.
\end{eqnarray*}
Now let us consider the conditions  
\begin{eqnarray*}
&&(X1)
\hspace{0.4cm}
\Phi(-+)> 0
\hspace{0.4cm}
{\rm and} 
\hspace{0.4cm}
\phi(-)> 0,
\\
&&(X2)
\hspace{0.4cm}
\Phi(-+)< 0
\hspace{0.4cm}
{\rm and} 
\hspace{0.4cm}
\phi(-)< 0,
\end{eqnarray*}
{\rm and} 
\begin{eqnarray*}
&&(Y1)
\hspace{0.4cm}
\Phi(+-)> 0
\hspace{0.4cm}
{\rm and} 
\hspace{0.4cm}
\phi(+)> 0,
\\
&&(Y2)
\hspace{0.4cm}
\Phi(+-)< 0
\hspace{0.4cm}
{\rm and} 
\hspace{0.4cm}
\phi(+)< 0.
\\
\end{eqnarray*}
The equations (\ref{x2y2cond}) are satisfied with real $x$ and $y$ if \begin{eqnarray}
((X1)\: {\rm or}\: (X2))
\hspace{0.4cm}
{\rm and} 
\hspace{0.4cm}
((Y1)\: {\rm or}\: (Y2)). 
\label{phasecond1}
\end{eqnarray}
Because of the fact that 
$\Phi(+-)> 0$ {\rm and} $\Phi(-+)< 0$ are equivalent, and that 
$\Phi(+-)< 0$ {\rm and} $\Phi(-+)> 0$ are equivalent, 
(\ref{phasecond1}) is equivalent to 
\begin{eqnarray}
((X1)\: {\rm and}\: (Y2))
\hspace{0.4cm}
{\rm or} 
\hspace{0.4cm}
((X2)\: {\rm and}\: (Y1)). 
\label{phasecond2}
\end{eqnarray}
Because of the fact that 
$\phi(-)> 0$ and $\phi(+)< 0$ yield $\Phi(-+)> 0$, and that 
$\phi(-)< 0$ and $\phi(+)> 0$ yield $\Phi(-+)< 0$,  
(\ref{phasecond2}) is equivalent to 
\begin{eqnarray}
(\phi(-)> 0\: {\rm and}\: \phi(+)< 0)
\hspace{0.3cm}
{\rm or} 
\hspace{0.3cm}
(\phi(-)< 0\: {\rm and}\: \phi(+)> 0). 
\label{phasecond3}
\end{eqnarray}
The condition (\ref{phasecond3}) determines the gapless region. 
From (\ref{ellipse1}) and (\ref{ellipse2}),
we find that all the boundaries of the gapless region determined by (\ref{phasecond3}) 
are gapless.  
We will here consider the following two cases: 
\\

\noindent
{\bf Case I.} $J_{x}\geq 0$, $J_{y}\geq 0$, $J_{z}\geq 0$, 
$J_{x}+J_{y}+J_{z}=1$, and $J_{4}\geq 0$.  
In this case (\ref{phasecond3}) is written as 
\begin{eqnarray*}
&&
(J_{x}-J_{y}-J_{4}+J_{z})(J_{x}-J_{y}+J_{4}-J_{z})> 0, 
\\
&&
\hspace{1.4cm}
{\rm and} 
\hspace{1.5cm}
J_{x}+J_{y}-J_{4}-J_{z}< 0,
\end{eqnarray*}
{\rm or}
\begin{eqnarray*}
&&
(J_{x}-J_{y}-J_{4}+J_{z})(J_{x}-J_{y}+J_{4}-J_{z})< 0, 
\\
&&
\hspace{1.4cm}
{\rm and} 
\hspace{1.5cm}
J_{x}+J_{y}-J_{4}-J_{z}> 0.
\end{eqnarray*}
When $J_{4}=0$, we find the phase diagram obtained by Kitaev\cite{06Kitaev}. 
\\


\noindent
{\bf Case II.} $J_{x}\geq 0$, $J_{y}\geq 0$, $J_{4}\geq 0$, 
$J_{z}=-J$, $J\geq 0$, and $J_{x}+J_{y}+J=1$. 
In this case, it can be derived that ((X1) and (Y2)) cannot be satisfied, 
and (\ref{phasecond3}) is written as 
\begin{eqnarray}
&&
(J_{x}-J_{y}-J_{4}-J)(J_{x}-J_{y}+J_{4}+J)< 0, 
\hspace{0.3cm}
\nonumber\\
&&
\hspace{0.6cm}
{\rm and} 
\hspace{0.5cm}
(1-J_{4})(J_{x}+J_{y}+J_{4}-J)> 0.
\hspace{0.3cm}
\label{KzeqnegK}
\end{eqnarray}
The condition  (\ref{KzeqnegK}) is satisfied only when $J_{4}< 1$. 


\section{Symmetries and Anyon excitations}\label{SymmetriesAndDiscussions}

\begin{figure}
\includegraphics{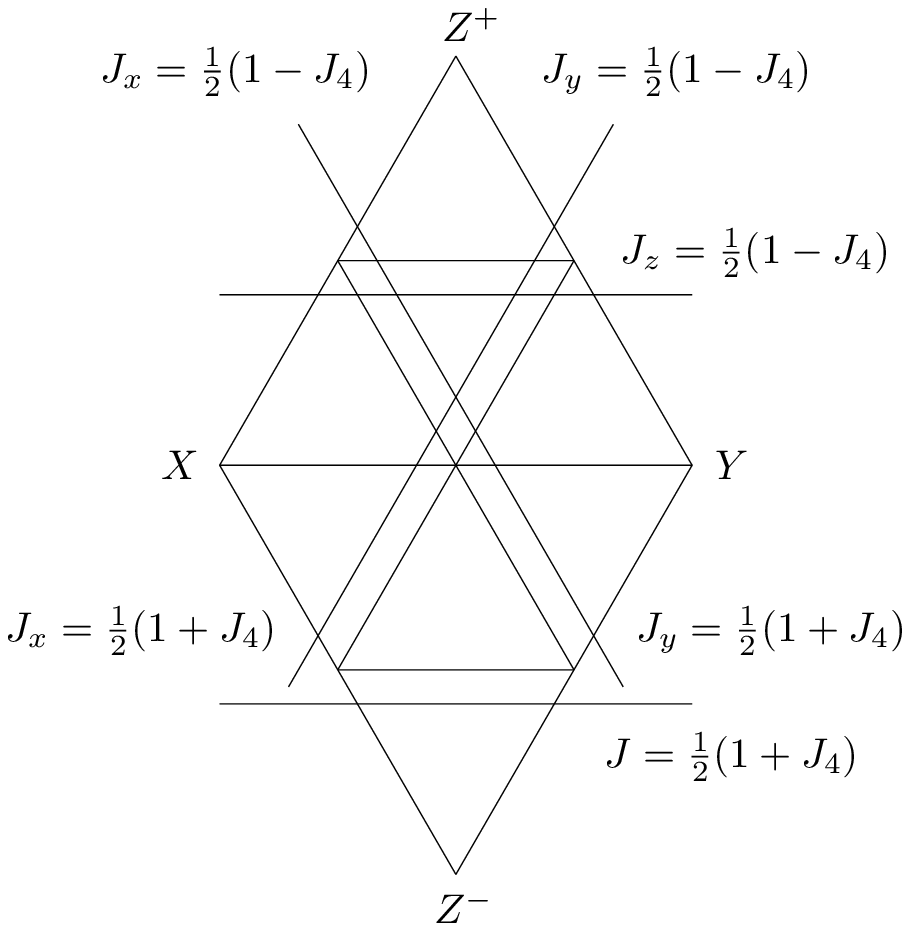}
\caption{\label{fig:PhaseDiagramGeneral} Four lines in Fig.10.}
\end{figure}

\begin{figure}
\includegraphics{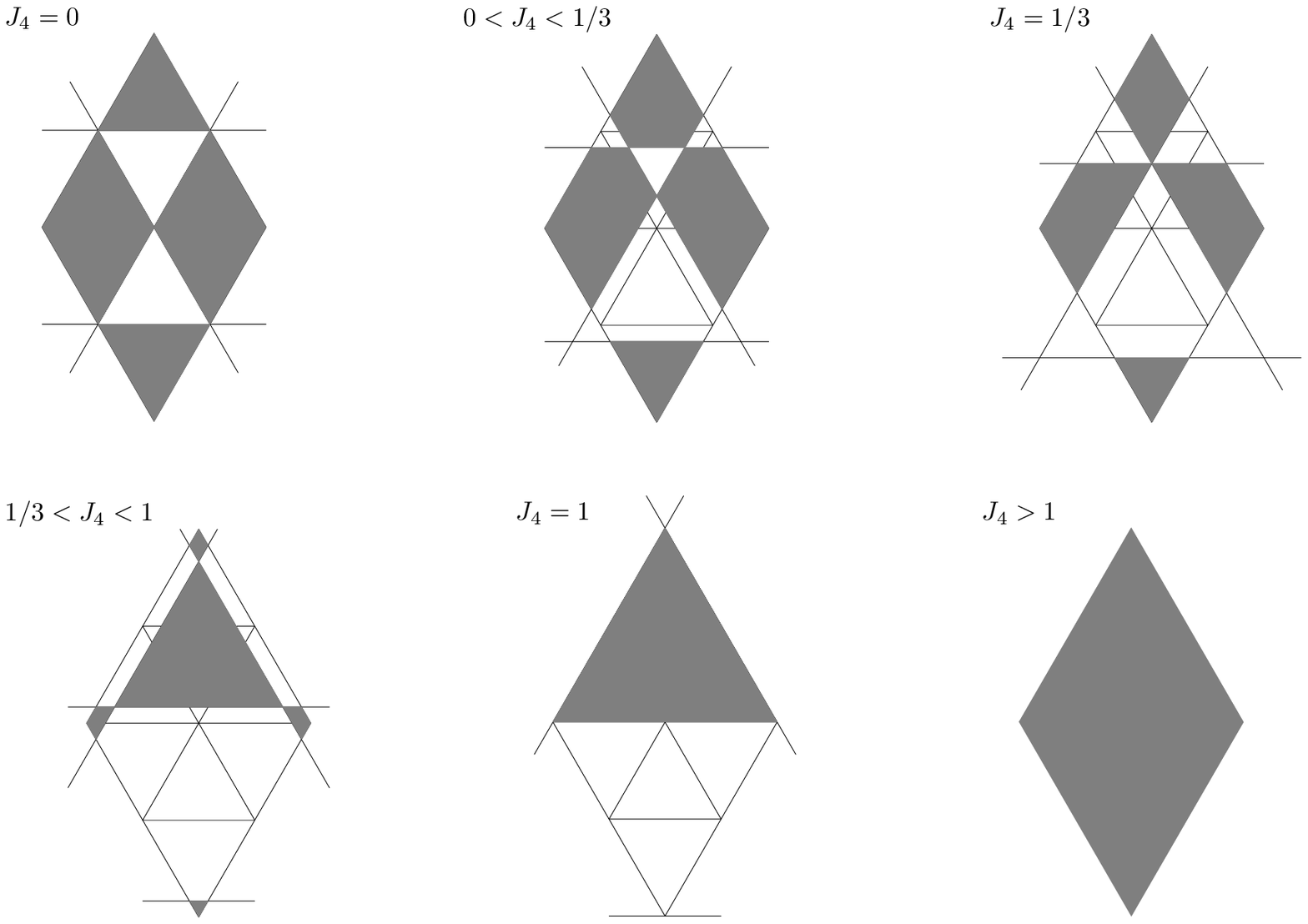}
\caption{\label{fig:PhaseDiagrams}The ground state phase diagram of the honeycomb lattice Kitaev model with Wen-Toric-code four-body interactions, where gapped regions are colored by gray.}
\end{figure}

The phase diagram is shown 
in Fig.\ref{fig:PhaseDiagramGeneral} and Fig.\ref{fig:PhaseDiagrams}. 
The triangle on the upper half plane corresponds to the case 
$J_{x}\geq 0$, $J_{y}\geq 0$, $J_{z}=J\geq 0$, and $J_{4}\geq 0$, 
and the triangle on the lower half plane corresponds to the case 
$J_{x}\geq 0$, $J_{y}\geq 0$, $J_{z}=-J\leq 0$, and $J_{4}\geq 0$.  
The interactions are normalized as $J_{x}+J_{y}+J=1$. 
Four corners are the points with the interactions 
$X$: $(J_{x}, J_{y}, J_{z})=(1, 0, 0)$, 
$Y$: $(J_{x}, J_{y}, J_{z})=(0, 1, 0)$, 
$Z_{+}$: $(J_{x}, J_{y}, J_{z})=(0, 0, 1)$, 
$Z_{-}$: $(J_{x}, J_{y}, J_{z})=(0, 0, -1)$. 
Two additional horizontal lines indicate 
$J_{z}=(1-J_{4})/2$ and $J=(1+J_{4})/2$.   
Other two lines indicate 
$J_{x}=(1-J_{4})/2$ $(J_{z}\geq 0)$, $J_{y}=(1+J_{4})/2$ $(J_{z}\leq 0)$, 
and 
$J_{y}=(1-J_{4})/2$ $(J_{z}\geq 0)$, $J_{x}=(1+J_{4})/2$ $(J_{z}\leq 0)$. 
The gapped regions are colored by gray.

Two ellipses (\ref{ellipse1}) and (\ref{ellipse2}) 
are invariant with the changes of signs 
(a) $(J_{x}, J_{y})\mapsto (-J_{x}, -J_{y})$ and/or 
(b) $(J_{4}, J_{z})\mapsto (-J_{4}, -J_{z})$. 
The phase diagram is thus invariant with these transformations. 
When we consider 
(c) $(J_{x}, J_{z})\mapsto (-J_{x}, -J_{z})$, 
(d) $(J_{y}, J_{z})\mapsto (-J_{y}, -J_{z})$, 
(e) $(J_{x}, J_{4})\mapsto (-J_{x}, -J_{4})$, 
(f) $(J_{y}, J_{4})\mapsto (-J_{y}, -J_{4})$, 
then each of (c)-(f) yields simultaneous interchange of $x$ and $y$ 
in (\ref{ellipse1}) and (\ref{ellipse2}).  
The phase diagram is thus still invariant with these transformations (c)-(f). 

These symmetries come from the symmetry of canonical rotations in spin space, 
and hence are also valid at finite temperatures. 
The invariance under (a)-(f) are explained 
from the invariance of the Hamiltonian, 
with the changes of the signs of interactions (a)-(f), 
together with the following canonical rotations, 
\begin{eqnarray*}
&(a)& (\sigma^{x}_{jl}, \sigma^{y}_{jl}, \sigma^{z}_{jl})\mapsto 
    (-\sigma^{x}_{jl}, -\sigma^{y}_{jl}, \sigma^{z}_{jl}) 
    \hspace{0.4cm}{\rm if}\hspace{0.4cm} j={\rm odd}, \\
&(b)& (\sigma^{x}_{jl}, \sigma^{y}_{jl}, \sigma^{z}_{jl})\mapsto 
    (\sigma^{x}_{jl}, -\sigma^{y}_{jl}, -\sigma^{z}_{jl}) 
    \hspace{0.4cm}{\rm if}\hspace{0.4cm} l={\rm odd}, \\
&(c)& (\sigma^{x}_{jl}, \sigma^{y}_{jl}, \sigma^{z}_{jl})\mapsto 
    (-\sigma^{x}_{jl}, \sigma^{y}_{jl}, -\sigma^{z}_{jl})
    \hspace{0.4cm}{\rm if}\hspace{0.4cm} j={\rm odd}, \\
&(d)& (\sigma^{x}_{jl}, \sigma^{y}_{jl}, \sigma^{z}_{jl})\mapsto 
    (\sigma^{x}_{jl}, -\sigma^{y}_{jl}, -\sigma^{z}_{jl}) 
    \hspace{0.4cm}{\rm if}\hspace{0.4cm} j={\rm odd}, \\
&(e)& (\sigma^{x}_{jl}, \sigma^{y}_{jl}, \sigma^{z}_{jl})\mapsto 
    (-\sigma^{x}_{jl}, \sigma^{y}_{jl}, -\sigma^{z}_{jl}) 
    \hspace{0.4cm}{\rm if}\hspace{0.4cm} j+l={\rm odd}, \\
&(f)& (\sigma^{x}_{jl}, \sigma^{y}_{jl}, \sigma^{z}_{jl})\mapsto 
    (\sigma^{x}_{jl}, -\sigma^{y}_{jl}, -\sigma^{z}_{jl}) 
    \hspace{0.4cm}{\rm if}\hspace{0.4cm} j+l={\rm odd}, \\
\end{eqnarray*}
respectively. 
As a result, the system is invariant 
changing the signs of arbitrary two interactions.  
The system with an even number of positive interactions 
are, therefore, equivalent to each other, 
and the system with an odd number of positive interactions 
are equivalent to each other. 
Thus the phase diagram given in Fig.\ref{fig:PhaseDiagrams}, 
with $J_{x}, J_{y}, J_{4}\geq 0$ 
and with $J_{z}\geq 0$ or $J_{z}\leq 0$, 
classifies all the possible cases. 

In case of $J_{4}=0$, the system is invariant 
changing the signs of $J_{x}$, $J_{y}$, and $J_{z}$, independently. 
The phase diagram, as a result, becomes symmetric, 
as shown in the first diagram in Fig\ref{fig:PhaseDiagrams}. 

Next let us consider the degeneracy of the ground state. 
In case of $(J_{x}, J_{y}, J_{z})=(J_{4}, (1-J_{4})/2, (1-J_{4})/2)$, 
two ellipses (\ref{ellipse1}) and (\ref{ellipse2}) become identical. 
In this case, for all $q_{1}$ there exists $q_{2}$ 
with which (\ref{eigenvalues}) becomes zero, 
and hence the ground state is highly degenerate.  
In case of $(J_{x}, J_{y}, J_{z})=((1-J_{4})/2, J_{4}, (1-J_{4})/2)$, 
two ellipses (\ref{ellipse1}) and (\ref{ellipse2}) become also identical, 
and in case of $(J_{x}, J_{y}, J_{z})=((1-J_{4})/2, (1-J_{4})/2, J_{4})$, 
both (\ref{ellipse1}) and (\ref{ellipse2}) 
become finite intervals on the real axis, 
and in these two cases, the ground state is also highly degenerate. 

When $J_{4}=1/3$, we have a symmetric point $J_{x}=J_{y}=J_{z}=J_{4}=1/3$, 
where above three points become identical, 
as shown in the third diagram in Fig\ref{fig:PhaseDiagrams}. 
In this case, (\ref{ellipse1}) and (\ref{ellipse2}) 
become finite intervals on the real axis. 
\\

When we consider the case with uniform $\Psi_{ij}$, 
we can find the ground state in this subspace, 
and because of the translational invariance, 
the Hamiltonian can be diagonalized in the momentum representation. 
In this subspace, 
the Hamiltonian is expressed symmetrically as the sum in (\ref{KxSym}) 
and sums coming from other interactions. 
Let us consider the replacement of the variables  
\begin{eqnarray}
(q_{1}, q_{2})\mapsto (-q_{1}, -q_{2}) 
\label{symmtrans1}
\end{eqnarray}
and accordingly 
\begin{eqnarray*}
&&
(J_{x}, J_{y})\mapsto (J_{y}, J_{x}), 
\hspace{0.3cm}
(J_{z}, J_{4})\mapsto (J_{4}, J_{z}), 
\nonumber\\
&&
c_{1}^{\dag}(-q_{1}, -q_{2})
=
c_{1}(q_{1}, q_{2})
\mapsto 
{\tilde c}_{1}^{\dag}(q_{1}, q_{2}),
\nonumber\\
&&
c_{2}(-q_{1}, -q_{2})
=
c_{2}^{\dag}(q_{1}, q_{2})
\mapsto 
{\tilde c}_{2}(q_{1}, q_{2}).
\end{eqnarray*}
Then the range of the summation 
$-\pi\leq q_{1}<\pi$ and $-\pi\leq q_{2}<\pi$ in (\ref{KxSym}) 
is invariant, 
and ${\tilde c}_{1}$ and ${\tilde c}_{2}$ 
satisfy the fermion anticommutation relations. 
We find from $h_{21}$ and $h_{12}$ in (\ref{hammatrixelements}) that the Hamiltonian 
${\cal H}(J_{x}, J_{y}, J_{z}, J_{4})$ 
and 
${\cal H}(J_{y}, J_{x}, J_{4}, J_{z})$ 
are equivalent. 
This is the particle-hole transformation. 

In this sense, the model with the interactions 
$J_{x}=J_{y}$, $J_{z}\neq 0$, $J_{4}=0$, 
and the model with 
$J_{x}=J_{y}$, $J_{z}=0$, $J_{4}\neq 0$ 
are equivalent. 

Kitaev\cite{06Kitaev} considered the large $J_{z}$ limit 
of the honeycomb lattice Kitaev model, 
and derived an effective Hamiltonian 
that consists of the Wen-type four-body interaction $J_{4}$ 
(see (37) in \cite{06Kitaev}). 
In this effective Hamiltonian, 
vortices are generated by two kinds of string operators, 
and an additional sign appears from each cross point of the strings 
when one interchange the positions of two excitations. 
In this sense the excitations are regarded as anyons.  

This fact is consistent with our argument 
that the large $J_{z}$ region is equivalent to the large $J_{4}$ region, 
and we thus also find that the abelian anyons appear 
in the large $J_{4}$ region as well as in the large $J_{z}$ region.

Let us again consider the replacement of the variables that  
\begin{eqnarray}
&&
(q_{1}, q)\mapsto (q, q_{1}) 
\hspace{0.3cm}
{\rm where} 
\hspace{0.3cm}
q=q_{1}/2+q_{2}, 
\label{symmtrans2}
\end{eqnarray}
and accordingly 
\begin{eqnarray*}
&&
(J_{x}, J_{y})\mapsto (J_{z}, J_{4}), 
\hspace{0.3cm}
(J_{z}, J_{4})\mapsto (J_{x}, J_{y}), 
\nonumber\\
&&
c_{1}^{\dag}(q_{1}, q_{2})
=
c_{1}^{\dag}(q_{1}, q-q_{1}/2)
\mapsto 
{\tilde c}_{1}^{\dag}(q, q_{1}),
\nonumber\\
&&
c_{2}(q_{1}, q_{2})
=
c_{2}(q_{1}, q-q_{1}/2)
\mapsto 
{\tilde c}_{2}(q, q_{1}). 
\end{eqnarray*}
It is easy to check that 
the summation over the region 
$-\pi\leq q_{1}<\pi$ and $-\pi\leq q_{2}<\pi$ 
is equivalent to the summation over 
$-\pi\leq q_{1}<\pi$ and $-\pi\leq q<\pi$, 
because of the periodic structure of the system with period $2\pi$. 
The operators  
${\tilde c}_{1}$ and ${\tilde c}_{2}$ 
satisfy the fermion anticommutation relations, 
and we find from (\ref{hammatrixelements}) that the Hamiltonian 
${\cal H}(J_{x}, J_{y}, J_{z}, J_{4})$ 
and 
${\cal H}(J_{z}, J_{4}, J_{x}, J_{y})$ 
are equivalent.  

In this sense, from (\ref{symmtrans1}) and (\ref{symmtrans2}), 
the model with the interactions 
$J_{y}=J_{z}=J_{4}$, $J_{x}=0$, 
and the model with 
$J_{x}=J_{z}=J_{4}$, $J_{y}=0$, 
are equivalent to the case 
$J_{x}=J_{y}=J_{z}$, $J_{4}=0$. 

In the original Kitaev model, 
in the gapless phases, 
an external field opens an energy gap, 
and the string operators generate vortices that behave as anyons.  
For the purpose to investigate this phenomena, 
let us consider the Fourier transformation 
in whole the Hilbert space. 
(Note that the Fourier transformation itself is always possible 
even if $\Psi_{jl}$ are not uniform, 
though the Hamiltonian ${\cal H}$ cannot be simplified 
in the subspace 
where ${\cal H}$ does not have translational invariance.)
It can be verified that the operators $\varphi_{\alpha}(j, l)$ are transformed as 
$\varphi_{\alpha}(j, l)\mapsto\varphi_{\alpha}(-j, -l)$ and 
$\varphi_{\alpha}(j, l)\mapsto\varphi_{\alpha}(j-l/2, l)$ 
with the transformations (\ref{symmtrans1}) and (\ref{symmtrans2}), 
respectively. 
So (\ref{symmtrans1}) and (\ref{symmtrans2}) 
correspond to change of locations in real space. 
The spin operators can be expressed by $\varphi_{\alpha}(j, l)$ as 
\begin{eqnarray*}
\sigma^{z}_{jl}
&=&
\eta_{2j-1\: l}
=
(+2i)\varphi_{2}(j, l)\varphi_{1}(j, l),
\\
\sigma^{x}_{jl}
&=&
\sqrt{2}
\Big(
\prod_{r=1}^{l-1}
\prod_{k=1}^{N}
\eta_{2k-1\: r}
\Big)
\Big(
\prod_{k=1}^{j-1}
\eta_{2k-1\: l}
\Big)
\varphi_{1}(j, l),
\\
\sigma^{y}_{jl}
&=&
\sqrt{2}
\Big(
\prod_{r=1}^{l-1}
\prod_{k=1}^{N}
\eta_{2k-1\: r}
\Big)
\Big(
\prod_{k=1}^{j-1}
\eta_{2k-1\: l}
\Big)
\varphi_{2}(j, l).
\end{eqnarray*}
The Zeeman term and the string operators are, therefore, 
transformed together with $\varphi_{\alpha}(j, l)$. 
In the subspace where $\Psi_{jl}$ are uniform, 
the Hamiltonian ${\cal H}$ is decomposed as (\ref{hammatrix}) 
according to the wave numbers. 
The Zeeman term and the string operators do not commute with $\Psi_{jl}$, 
thus they are not simple in this momentum bases, 
they change their locations, and generate anyons. 

\section{Conclusion}\label{Conclusion}

At last we would like to note an interesting methodology  
presented in \cite{09Nussinov} and \cite{11Cobanera}, 
in which isomorphisms of algebras 
that are generated from interactions 
are considered, 
and equivalences and mappings are investigated. 
In \cite{09Nussinov}, results of \cite{06Kitaev} and \cite{08ChenNussinov} 
on the honeycomb lattice Kitaev model 
was rederived by the algebraic isomorphism, 
and  in \cite{11Cobanera}, the Jordan-Wigner transformation is generated 
in an iterative way in the case of the XY chain. 
The basic idea in these papers 
that the algebraic structure of interactions determine the spectrum of the model 
is common to our formula. 
In the present paper, however,  
the transformation (\ref{transmain}) is explicitly given 
for the series of operators that satisfy (\ref{cond}), 
and the two-dimensional systems are investigated.

In summary, we obtain the exact ground state phase diagram 
of the honeycomb lattice Kitaev model 
with the Wen-Toric-code four-body interactions, 
and find that the structure of the system is symmetric 
in four-dimensional space $(J_{x}, J_{y}, J_{z}, J_{4})$. 
The fermionization transformation (\ref{transmain}) 
is generally formulated for two-dimensional systems. 
The construction of the series of operators that satisfy (\ref{cond}) 
is also generalized, 
and a model equivalent to the Kitaev model is introduced. 
We also find that the anyon excitations appear 
in all of the regions shown in the phase diagram,  
they can be transformed each other. 
\\




\begin{thebibliography}{00}


\bibitem{16Minami}
K. Minami, 
J.~Phys.~Soc.~Jpn. 85, 024003 (2016). 

\bibitem{17Minami}
K. Minami, 
Nucl.~Phys.~B~925, 144 (2017).

\bibitem{18Yanagihara}
Y. Yanagihara and K. Minami, 
in preparation.

\bibitem{71Suzuki} 
M. Suzuki, 
Prog. Theor. Phys. 46, 1337 (1971).

\bibitem{06Kitaev}
A. Kitaev, 
Ann. Phys. 321, 2 (2006).


\bibitem{07LeeZhangXiang}
D-H. Lee, G-M. Zhang, and T. Xiang, 
Phys. Rev. Lett. 99, 196805 (2007).

\bibitem{09ShiYuYouNori}
X-F. Shi, Y. Yu, J. Q. You, and F. Nori, 
Phys. Rev. B~79, 134431 (2009).

\bibitem{08YuWang}
Y. Yu, and Z. Wang, 
Euro Phys. Lett. 84, 57002 (2008).

\bibitem{08Yu}
Y. Yu, 
Nucl. Phys. B~799, 345 (2008).

\bibitem{03Wen}
X-G. Wen, 
Phys. Rev. Lett. 90, 016803 (2003).

\bibitem{03Kitaev}
A. Kitaev, 
Ann. Phys. 303, 2 (2003).


\bibitem{07Feng}
X-Y. Feng, G-M. Zhang, and T. Xiang, 
Phys. Rev. Lett. 98, 087204 (2007).

\bibitem{07ChenHu}
H-D. Chen, and J. Hu, 
Phys. Rev. B~76, 193101 (2007).

\bibitem{08ChenNussinov}
H-D. Chen, and Z. Nussinov, 
J. Phys. A: Math. Theor. 41, 075001 (2008).



\bibitem{94Lieb}
E. H. Lieb, 
Phys. Rev. Lett. 73, 2158 (1994).



\bibitem{09Nussinov}
Z. Nussinov, and G. Ortiz,
Phys.~Rev. B 79, 214440 (2009).

\bibitem{11Cobanera}
E. Cobanera, G. Ortiz, and Z. Nussinov, 
Adv.~Phys. 60, 679 (2011).

\end{thebibliography}



\newpage

\setlength{\evensidemargin}{-1.2cm}
\setlength{\oddsidemargin}{-1.2cm}

\begin{table}
\begin{tabular}{llcc}
Fermi operators 
&
original operators 
&
series (\ref{seriesTrIsing})
&
series (\ref{seriesClusterNnn})
\\
\hline
$(+2i)\:\varphi_{2}(j, l)\varphi_{1}(j+2, l)$
&
$\eta_{2j\:  l}\eta_{2j+1\:  l}\eta_{2j+2\:  l}$
&
$-\sigma^{x}_{j\: l}\sigma^{z}_{j+1\: l}\sigma^{x}_{j+2\: l}$
&
$-\sigma^{x}_{2j\: l}\sigma^{x}_{2j+1\: l}\sigma^{z}_{2j+2\: l}\sigma^{x}_{2j+3\: l}\sigma^{x}_{2j+4\: l}$
\\
$(-2i)\:\varphi_{2}(j, l)\varphi_{1}(j+1, l)$
&
$\eta_{2j\:  l}$
&
$\sigma^{x}_{j\: l}\sigma^{x}_{j+1\: l}$
&
$\sigma^{x}_{2j\: l}1_{2j+1\: l}\sigma^{x}_{2j+2\: l}$
\\
$(+2i)\:\varphi_{2}(j, l)\varphi_{1}(j, l)$
&
$\eta_{2j-1\: l}$
&
$\sigma^{z}_{j\: l}$
&
$\sigma^{x}_{2j-1\: l}\sigma^{z}_{2j\: l}\sigma^{x}_{2j+1\: l}$
\\
$(-2i)\:\varphi_{2}(j, l)\varphi_{1}(j-1, l)$
&
$\eta_{2j-3\: l}\eta_{2j-2\: l}\eta_{2j-1\: l}$
&
$\sigma^{y}_{j-1\: l}\sigma^{y}_{j\: l}$
&
$\sigma^{x}_{2j-3\: l}\sigma^{y}_{2j-2\: l}1_{2j-1\: l}\sigma^{y}_{2j\:l}\sigma^{x}_{2j+1\: l}$
\\
$(+2i)\:\varphi_{2}(j, l)\varphi_{1}(j-2, l)$
&
$\eta_{2j-5\: l}\eta_{2j-4\: l}\eta_{2j-3\: l}\eta_{2j-2\: l}\eta_{2j-1\: l}$
&
$-\sigma^{y}_{j-2\: l}\sigma^{z}_{j-1\: l}\sigma^{y}_{j\: l}$
&

\\
\hline
$(+2i)\:\varphi_{2}(j, l)\varphi_{2}(j+2, l)$
&
$(+i)\eta_{2j\: l}\eta_{2j+1\: l}\eta_{2j+2\: l}\eta_{2j+3\: l}$
&
$-\sigma^{x}_{j\: l}\sigma^{z}_{j+1\: l}\sigma^{y}_{j+2\: l}$
&

\\
$(-2i)\:\varphi_{2}(j, l)\varphi_{2}(j+1, l)$
&
$(+i)\eta_{2j\:  l}\eta_{2j+1\:  l}$
&
$\sigma^{x}_{j\: l}\sigma^{y}_{j+1\: l}$
&
$\sigma^{x}_{2j\: l}\sigma^{x}_{2j+1\: l}\sigma^{y}_{2j+2\: l}\sigma^{x}_{2j+3\: l}$
\\
\hline
$(+2i)\:\varphi_{1}(j, l)\varphi_{1}(j+1, l)$
&
$(-i)\eta_{2j-1\: l}\eta_{2j\: l}$
&
$\sigma^{y}_{jl}\sigma^{x}_{j+1\: l}$
&
$\sigma^{x}_{2j-1\: l}\sigma^{y}_{2j\: l}\sigma^{x}_{2j+1\: l}\sigma^{x}_{2j+2\: l}$
\\
$(-2i)\:\varphi_{1}(j, l)\varphi_{1}(j+2, l)$
&
$(-i)\eta_{2j-1\: l}\eta_{2j l}\eta_{2j+1\: l}\eta_{2j+2\: l}$
&
$-\sigma^{y}_{j\: l}\sigma^{z}_{j+1\: l}\sigma^{x}_{j+2\: l}$
&
\end{tabular}
\caption{\label{tab:table3}
Relations between $\varphi_{\alpha}(j, l)$, 
$\eta_{j}$, and $\sigma^{k}_{jl}$,   
obtained by (\ref{transmain}) and (\ref{newnotation}).  
Operators $\eta_{j}$ are defined 
in (\ref{seriesTrIsing}) and in (\ref{seriesClusterNnn}). 
In the case of the series (\ref{seriesTrIsing}), for example,  
$(+2i)\:\varphi_{2}(j, l)\varphi_{1}(j, l)
=\eta_{2j-1\: l}
=\sigma^{z}_{jl}$, 
and 
in the case of the series (\ref{seriesClusterNnn}), 
$(+2i)\:\varphi_{2}(j, l)\varphi_{1}(j, l)
=\eta_{2j-1\: l}
=\sigma^{x}_{2j-1\: l}\sigma^{z}_{2j\: l}\sigma^{x}_{2j+1\: l}$.
}
\end{table}


\vspace{3.0cm}

\newpage

\end{document}